\documentclass[submission, Phys]{SciPost}

\usepackage{lineno}

\usepackage{graphicx}

\usepackage{bm}
\usepackage{amsmath}
\usepackage{amsfonts}
\usepackage{amssymb}
\usepackage{color}
\usepackage{sectsty}
\usepackage{dsfont}
\usepackage{caption}
\usepackage{subcaption}
\usepackage{units}
\usepackage[normalem]{ulem}

\newcommand{\ket}[1]{\left|#1\right>}

\usepackage{xcolor}

\newcommand{\be}{\begin{equation}}
\newcommand{\ee}{\end{equation}}
\newcommand{\bi}{\begin{itemize}}
\newcommand{\ei}{\end{itemize}}

\begin{document}

\begin{center}{\Large \textbf{
Many-body localization in presence of cavity mediated long-range interactions
}}\end{center}

\begin{center}
Piotr Sierant$^1$,  Krzysztof Biedro\'n$^1$, Giovanna Morigi$^2$, and Jakub Zakrzewski$^{1,3}$
\end{center}

\begin{center}
{\bf 1} Instytut Fizyki imienia Mariana Smoluchowskiego, Uniwersytet Jagiello\'nski, \L{}ojasiewicza 11, 30-348 Krak\'ow, Poland
\\
{\bf 2} Theoretische Physik, Universit\"at des Saarlandes, D-66123  Saarbr\"ucken, Germany
\\
{\bf 3} Mark Kac Complex Systems Research Center, Jagiellonian University, \L{}ojasiewicza 11, 30-348 Krak\'ow, Poland
\\
* piotr.sierant@uj.edu.pl
\end{center}

\begin{center}
\today
\end{center}


\section*{Abstract}
{\bf
We show that a one-dimensional Hubbard model with all-to-all 
coupling may exhibit many-body localization in the presence of
local  disorder. We numerically identify the parameter space where 
many-body localization occurs using exact diagonalization and 
finite-size scaling. The time evolution from a random initial 
state  exhibits 
features consistent with the localization picture. The dynamics 
can be observed with quantum gases in optical cavities, localization 
can be revealed through the time-dependent dynamics of the light emitted by the resonator.
}

\vspace{10pt}
\noindent\rule{\textwidth}{1pt}
\tableofcontents\thispagestyle{fancy}
\noindent\rule{\textwidth}{1pt}
\vspace{10pt}

\section{Introduction}
\label{sec:intro}

Many-body localization (MBL) is the most robust manifestation of ergodicity 
  breaking in interacting many-body systems. The interactions are generically considered as leading 
  to ergodic dynamics as far as 
local observables are concerned. 
The standard formulation of this belief is the eigenvector 
thermalization hypothesis (ETH) \cite{Deutsch91,Srednicki94}. Numerous studies over the last decade
have shown that the many-body interacting systems do
not thermalize in presence of a quenched disorder. {In particular, for a sufficiently strong disorder MBL may occur, leading to a long-time
memory of the initial state} (for recent 
reviews see \cite{Nandkishore15} as well as a topical issue
of Annalen der Physik \cite{AnnPhys:MBL:2017}).

While most studies so far considered short-range interactions, for instance in the XXZ spin Hamiltonian 
(which became a paradigmatic model of MBL \cite{Pal10}) or bosons and fermions in 
optical lattices in {the tight--binding limit} \cite{Mondaini15,Schreiber15,Choi16,Sierant17,Sierant18},
it is by no means clear whether MBL persists for genuinely long-range interactions, 
such as for Coulombic or dipolar potentials.
In recent works \cite{Nandkishore17} it was argued that MBL  may appear in disordered 
long-range interacting systems, others suggest
 \cite{Burin15,Li16,Ho18,Tikhonov18} lack of MBL for, e.g., dipole-dipole interactions in
 three dimensions (3D) \cite{Levitov90,Yao14}.  
Furthermore, MBL in presence of power-law decaying tunneling elements and interactions has
been studied in
\cite{Safavi18,DeTomasi18,Botzung18} where algebraic decay of correlation functions as 
well as of algebraic growth of entanglement entropy was found. In other studies
\cite{Ponte2017,Hetterich2017,Hetterich2018,Ng18} a model of a single spin coupled to all sites 
of otherwise short-range interacting systems was analysed showing that the presence of 
additional coupling may strongly modify MBL properties.

In the present work we numerically analyse whether MBL occurs in a disordered Hubbard model 
with all-to-all interactions.
This model is expected to describe the dynamics 
of atoms in an external lattice and interacting dispersively with a mode
of a standing-wave optical cavity. Such a model has been extensively studied in the past 
concentrating mainly, however, on ground state properties
\cite{Mekhov07,Maschler08,Larson08,Fernandez10,Leroux2010,Mueller12,Mottl12,
Habibian13,Habibian13b,Dogra16,Niederle16,Rojan16,Landig16} {(for a review with 
extensive list of references to earlier works see \cite{Ritsch13}. The certainly  
incomplete list of recent experimental works in that area includes \cite{Georges2018,Davis2019,Braverman2019,Morales2019})}. Long-range interactions appear naturally 
in this system -- the mode of the cavity mediates a two-body interaction whose 
range is as large as the system size. {When the atoms are tightly confined by an external optical lattice, the cavity-mediated long-range interactions tends to order the atoms in structures maximizing the intracavity field intensity. We investigate MBL in this extended Hubbard model and with local disorder} using exact 
diagonalization supplemented by numerical techniques for 
sparse Hamiltonian matrices for a {gas} of fermions (or bosons).
Our results show features which can be attributed to the occurence of MBL in the system. {We argue that these features can be revealed in the light emitted by the resonator}.
 
The paper is structured as follows. In Sec.~\ref{description} we describe an experimentally 
realizable system of atoms inside a resonant cavity. {We summarize the details of the derivation of the effective model in Appendix A,  as described e.g. in
\cite{Fernandez10,Habibian13b}}. In Sec.~\ref{phasediagram}
we determine the phase diagram by a finite-size scaling analysis assuming that the atoms are
spinless fermions. In Sec.~\ref{correlations} we 
 turn to dynamics of the system and provide the evidence of ergodicity breaking
{resulting from the interplay of disorder and interactions in the system}. 
In Sec.~\ref{weakLIOMs} we show that the ergodicity breaking can be understood within an appropriately 
modified picture of local integrals of motion (LIOMs). We then 
argue that in cavity QED setups ergodicity breaking can be revealed via the light emitted at the cavity output.
 In 
Sec.~\ref{quasi} we provide arguments that the reported properties on the system are similar 
both for purely random disorder as well as 
for quasiperiodic potential. Finally, we discuss MBL for bosons loaded into the cavity system. 
{Appendix B contains a short discussion of the 
nonergodicity in the specific limit of very strong cavity-mediated coupling which is outside the MBL regime; Appendix C provides the 
details of the time-propagation algorithm used.}

\section{Description of the system}
\label{description}

We consider an ensemble of atoms trapped in a quasi {one-dimensional} geometry and tightly bound by an optical lattice.
The atoms dispersively interact with an optical cavity in the regime, in which the cavity mode can be 
adiabatically eliminated from the dynamics. We assume that the cavity field is a perturbation to the optical 
lattice, so that the dynamics can be restricted to the lowest band of the external lattice and
the cavity-mediated long-range interactions effectively
describe all-to-all interactions between the lattice sites. The dynamics is governed by the effective Hamiltonian (for the detailed derivation see Appendix A):
\begin{equation}
\label{H1} 
H=H_{\rm A}+H_{\rm C}\,, 
\end{equation}
 \begin{figure}[htbp]
\begin{center}
\includegraphics[scale=0.55]{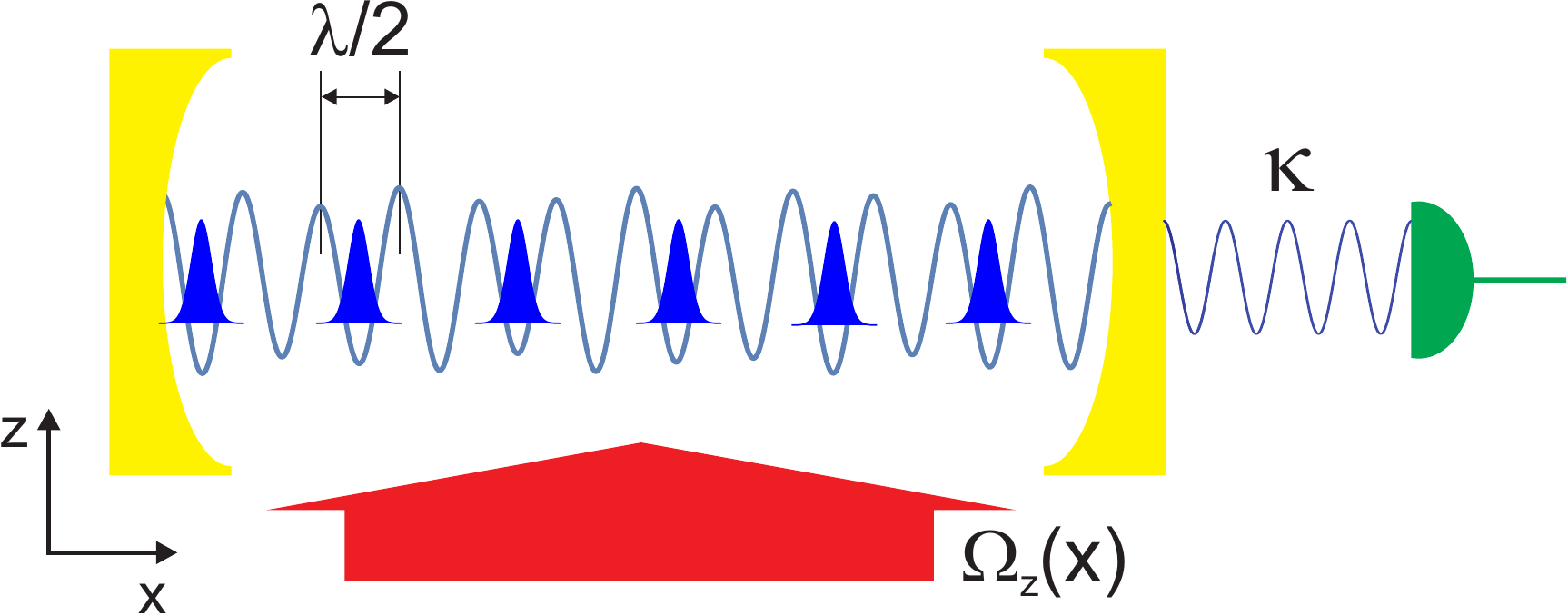}
\caption{(color online) The Hubbard model we consider  describes the dynamics of atoms in an optical 
lattice and interacting with the standing-wave mode of a high-finesse optical cavity. In the limit in 
which the atom-photon interactions are dispersive and the cavity field can be adiabatically eliminated 
from the atomic dynamics, the resulting Hubbard model is characterized by all-to-all interactions 
mediated by the cavity photons. The trasverse arrow symbolizes an external laser pumping photons into 
the cavity via coherent atom scattering. The additional disorder 
shifts the energy of the lattice potential. The quantum state of the system can be inferred by measuring the 
light at the cavity output, by time-of-flight measurements, 
or by Bragg spectroscopy using a weak probe. }
\label{scheme}
\end{center}
\end{figure}
with  $H_{\rm A}$ being the  standard Hubbard-like Hamiltonian for 
the dynamics of atoms in an optical lattice and in presence of disorder
and $H_{\rm C}$ the energy of the interaction with the cavity field. In detail,
the optical lattice is composed of $K$ sites, we assume 
periodic boundary conditions 
and the atomic Hamiltonian reads
\begin{equation}
\label{H0}
H_{\rm A}=-J\sum_j^{K}\left(b_{j+1}^\dagger b_j+{\rm H.c.}\right)+\sum_j^K E_j n_j +H^{\rm F,B}_{\rm int}\,.
\end{equation}
where $b_j$ and $b_j^\dagger$ are the onsite annihilation and creation operators of a fermion 
or a boson at site $j=1,\ldots,K$, (with $K+1$ identified with the first site), $J$ is the tunneling coefficient scaling the nearest-neighbour hopping, 
$n_j=b_{j}^\dagger b_j$ is the occupation operator at site $j$, $E_j$ is the onsite energy
at site 
$j$, and $H^{\rm F,B}_{\rm int}$ is the interaction term, 
which takes different forms depending on the quantum statistics of the atomic gas. For spinless fermions
 \begin{equation}
\label{Hf} 
H^{\rm F}_{\textrm {int }}=U\sum_j^{K} n_j n_{j+1},
\end{equation}
and $U>0$, whereas the first non--trivial interaction term in tight-binding expansion for bosons reads
\begin{equation}
\label{Hb} 
H^{\rm B}_{\textrm {int }}=U\sum_j^{K} n_j (n_{j}-1).
\end{equation}
In turn, the cavity-mediated long-range interactions take the form \cite{Fernandez10,Habibian13b,Landig16}:
\begin{equation}
\label{H_c}
H_{\rm C}=- \frac{U_1}{K} \left(\sum_j^{K} (-1)^j n_{j} \right)^2 =- \frac{U_1}{K} \sum_{i,j}^{K} (-1)^{i+j} n_{i}n_{j} \,, 
\end{equation}
with $U_1>0$. This {Hamiltonian term} is here derived under the
assumption that the wavelength of the cavity field equals 
the one of the electric field generating the optical lattice, see Appendix A. 
{This interaction is proportional to the squared population imbalance, $H_{\rm C}\propto -I(t)^2$, 
where
\begin{equation}
\label{Eq:I}
 I(t)=  \sum_i (-1)^{i} n_i
\end{equation} 
and the sign is due to the fact that the standing wave cavity mode with wave number $k$ takes value 
$\cos (kia)=(- 1)^i$ at the optical lattice site centered at $x_i=ia$.
Its expectation value is thus positive (negative) when the even (odd) sites of the cavity standing-wave mode are prevailingly occupied and it favours} density-wave (DW) ordering \cite{Fernandez10,Habibian13b,Dogra16,Niederle16}.
We determine the existence of the MBL phase by an exact diagonalization of the Hubbard
Hamiltonian, {Eq.~\eqref{H1}, for spinless fermions}. This situation is 
in fact more accessible to numerical analysis since the local Hilbert 
space has only dimension $2$. We then only briefly show that analogous 
properties of the fermionic case are also found for bosons.  

We finally note that the disorder in our model is in the onsite energy. Here we assume two cases. Throughout most of the work we make the theoretically elegant 
assumption that $E_j$ are uncorrelated  random variables uniformly distributed in $[-W,W]$ interval, {where $2W$ denotes the interval width}. 
In Sec. \ref{quasi} we then analyse the situation where $E_j$ is due to a quasi-periodic optical potential.

In the rest of this manuscript we report energies in units of $J$ and time in units of $1/J$.

\section{Phase diagram}
\label{phasediagram}

Energy level statistics encode an answer to the question of whether a disordered system is 
localized or ergodic and satisfies ETH. Level statistics of ergodic systems with (generalized) time reversal symmetry have properties akin to the Gaussian Orthogonal Ensembe (GOE)
\cite{Haake}.  As the disorder strength increases and the system
becomes localized, the level statistics becomes Poissonian \cite{Serbyn16, Bertrand16} (an accurate
model for level statistics across the {localization} transition was recently proposed in Refs. 
\cite{Sierant18b, Sierant18c}).
The level statistics can be characterized using the gap ratio. This is defined as 
\begin{equation}
r_n = \mathrm{min}(\delta_n,\delta_{n+1})/ \mathrm{max}(\delta_n,\delta_{n+1}) 
\label{r:n}
\end{equation}
 with $\delta_n=E_n-E_{n-1}$ being the spacing 
between two consecutive eigenvalues \cite{Oganesyan07}. Averaging over different energy 
levels within a certain interval as well as over disorder realizations results in the mean 
gap ratio, $\overline r$,
that may be used to characterize the spectra. 
The mean gap ratio
changes from $\overline r\approx 0.53$ in the ergodic regime \cite{Oganesyan07,Atas13} to 
$\overline r\approx 0.39$ for a localized system and is thus a straightforward probe of the 
MBL transition 
{especially as it does not require level unfolding, a tricky procedure for a many body system \cite{Gomez02}.}

\begin{figure}[!ht]
\begin{center}
\includegraphics[width=0.8\columnwidth]{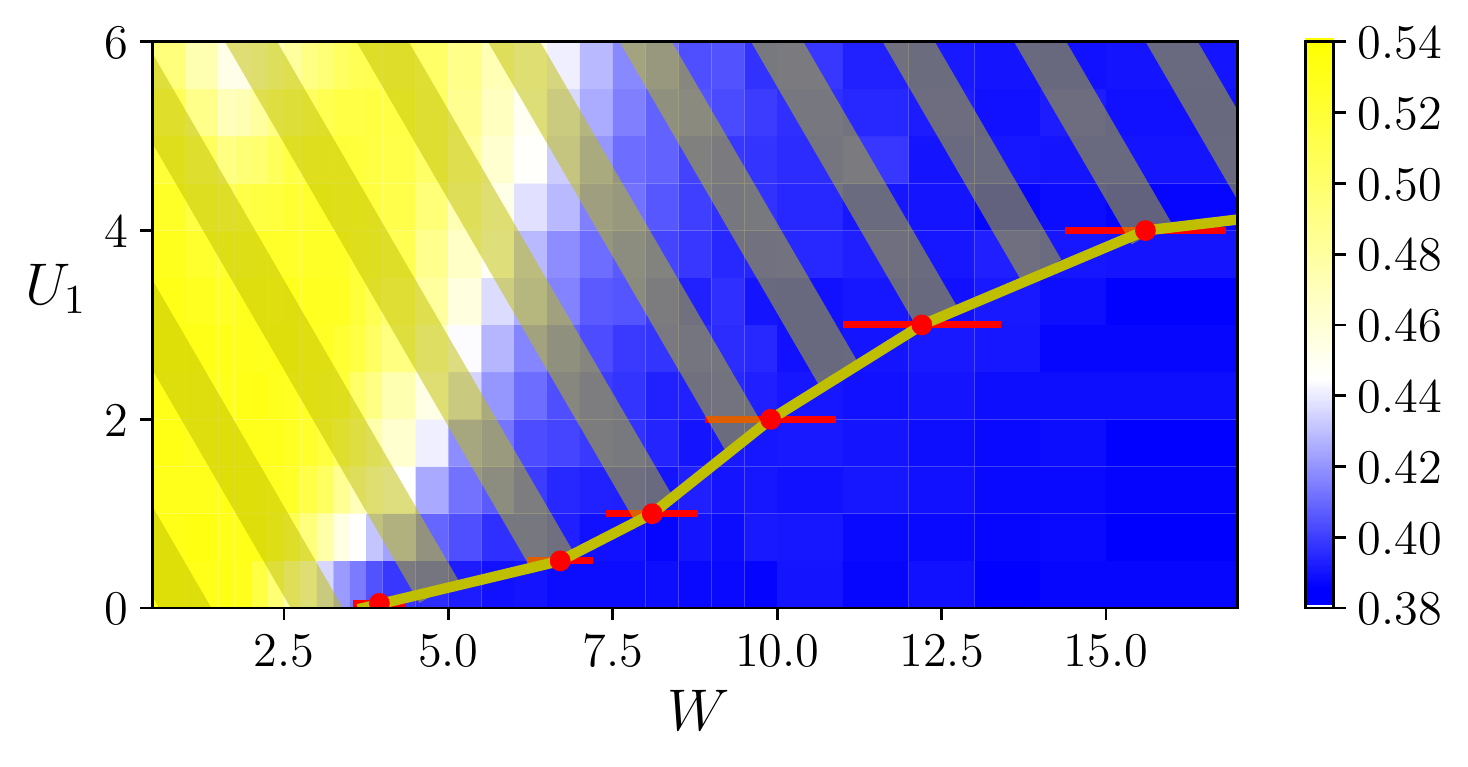}
\caption{Contour plot of the mean gap ratio $\overline r$ as a function of random
uniform disorder with amplitude $W$ 
and of the long-range interaction strength $U_1$. The mean gap ratio is determined in 
the center of the spectrum 
(for $\epsilon_n = (E_n - E_{\rm min})/(E_{\rm max} -E_{\rm min})\approx 0.5$). The yellow
and blue regions denote the ETH and the MBL phase, respectively, for $N=8$ fermions, $K=16$ 
lattice sites, and
 short-range interaction strength $U=1$. A finite-size scaling analysis (see text and Fig.~\ref{scaling}) 
 places the boundary between ETH and MBL at the 
solid yellow line connecting the markers, giving the critical disorder amplitude $W_C( U_1)$. 
The error bars {reported on the markers} result from the 
comparison of finite-size scaling {studies}
performed for $K=16,18,20$ and $K=14, 16,18,20$.  The statistical errors for the color-map data
points are well below 1\% of their values.
}
\label{rstat}
\end{center}
\end{figure}

Figure~\ref{rstat} displays the contour plot of the mean gap ratio $\overline r$ in the 
$W-U_1$ phase diagram, namely, as a function of the disorder and of the long-range interaction 
strength. The colour code refers to the calculations performed for 
a gas of $N=8$ fermions in a lattice with $K=16$ sites, the gap ratio was first determined
{for}
$500$ eigenvalues $E_n$ for which  $\epsilon_n = (E_n - E_{\rm min})/(E_{\rm max} -E_{\rm min})\approx 0.5$, 
where
$E_{\rm max},$ $E_{\rm min}$ are respectively the largest and the smallest eigenvalue for given 
disorder realization, and then averaged over $400$ disorder
realizations. {The statistics is sufficient to determine the mean gap ratio with an accuracy below 1\% of its value.}
One clearly identifies two regions: (i) the yellow region, corresponding 
to $\overline r \approx 0.53$ where the system has GOE level statistics and
is thus ergodic, and (ii) the blue region with
$\overline r \approx 0.39$, where the system is MBL. The white stripe separates
the ETH from the MBL regimes and gives the  disorder strength 
 at
which $\overline r = 0.45$. This {disorder} strength depends on the system size and
{shifts} to larger values as we increase {the system size} $K$. 

\begin{figure}[!ht]
\begin{center}
\includegraphics[width=0.5\columnwidth]{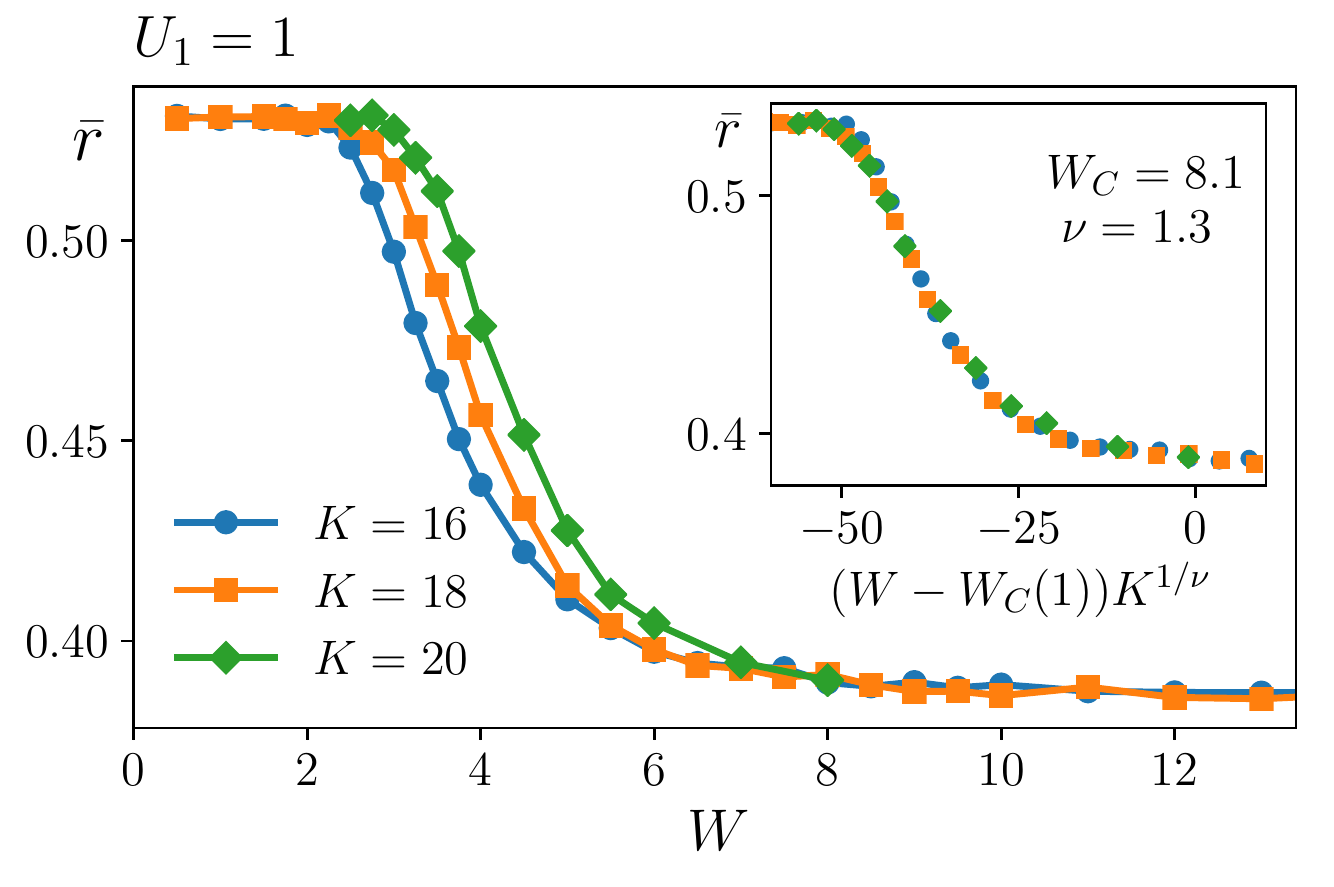}\includegraphics[width=0.5\columnwidth]{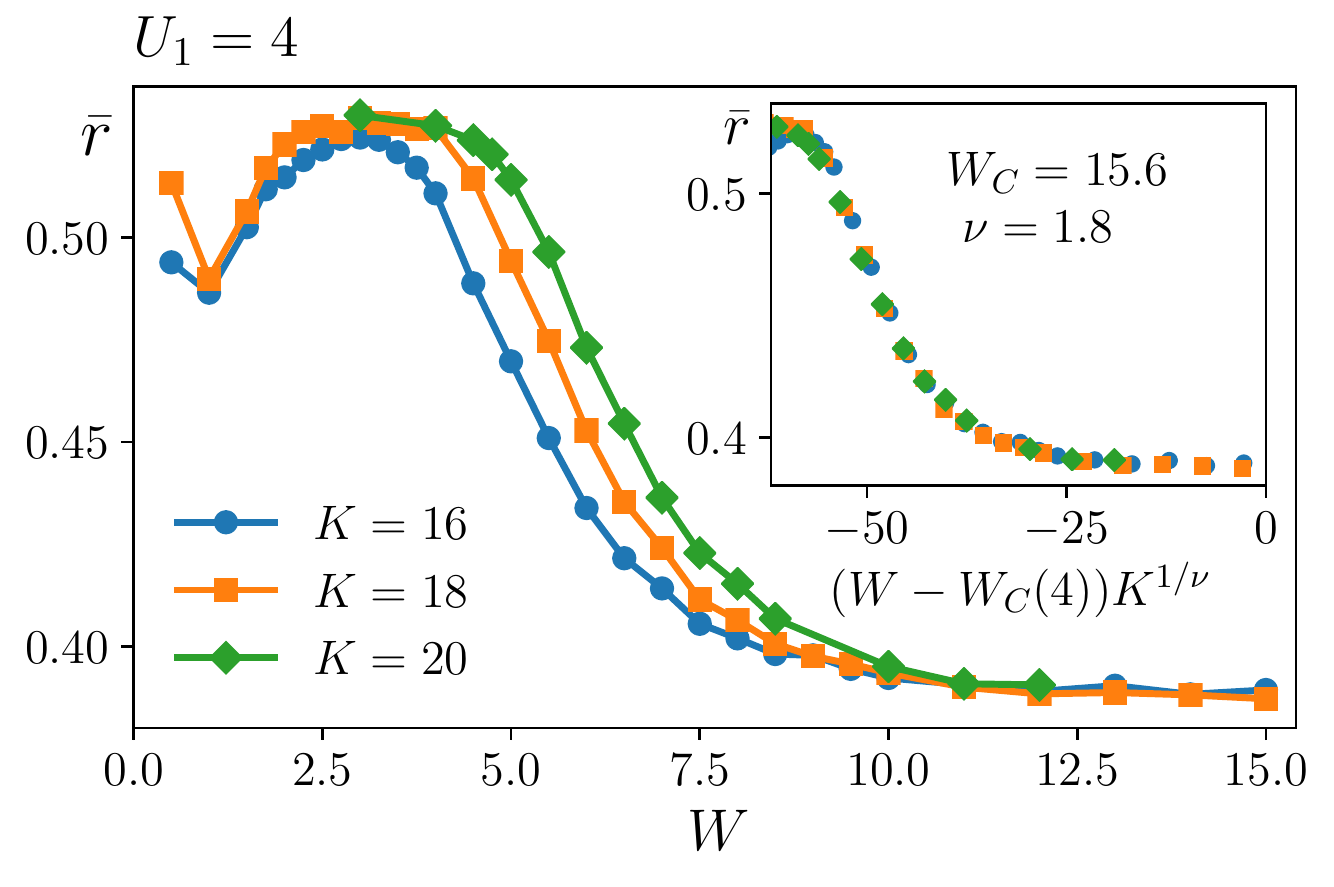}
\caption{Onset: The mean gap ratio $\overline r$, in the center of the spectrum, is displayed as a function of the disorder amplitude $W$ and for $K=16,18,20$ lattice sites. 
The left panel corresponds to $U_1=1$ , the right one to $U_1=4$. 
The insets display the data rescaled according to Eq.~\eqref{eqscal}.
Error bars are smaller than the marker's size. 
The universal functions $g_{U_1}[\left(W-W_C(U_1)\right)K^{1/\nu} ]$ are modeled by third order polynomials, points 
 with $\overline r \in [0.392, 0.48]$ are taken into account in the finite-size scaling procedure. }
\label{scaling}
\end{center}
\end{figure}

Nevertheless, 
a finite-size scaling
analysis suggests that {the ergodic-MBL boundary}
converges to the yellow line connecting the red markers.  
The red markers are obtained as follows. 
We first consider the scaling form of the disorder strength given by
\begin{equation}
 W \rightarrow \left(W-W_C(U_1) \right)K^{1/\nu(U_1)},
 \label{eqscal}
\end{equation}
where the critical disorder strength $W_C(U_1)$ and the exponent $\nu(U_1)$ 
depend on the long-range interaction strength $U_1$. 
{We then consider the system sizes $K=16,18,20$, which can be numerically simulated using the shift-invert 
technique, Ref. 
\cite{Pietracaprina18}, implemented in Portable, Extensible Toolkit for Scientific Computation (PETSc) in Scalable Library for Eigenvalue Problem computations (SLEPc) setting, see Refs. \cite{Balay18,Hernandez05}. The onset of Fig.~\ref{scaling} 
displays the mean gap ratio $\bar r$ in the band center as a function of disorder strength for
$U_1=1$ and $U_1=4$.} The scaling \eqref{eqscal} allows us to collapse the mean 
gap ratio for different system sizes $\overline r$ as a function of disorder strength $W$ onto universal curves 
{$g_{U_1}[\left(W-W_C(U_1)\right)K^{1/\nu} ]$ }
with good accuracy. From these curves we extract the critical disorder strengths $W_C(U_1)$ for 
$U_1 \in \{ 0, 0.5, 1, 2, 3, 4\}$, which 
correspond to the markers in Fig.~\ref{rstat}. 
From this ansatz we also determine the exponent $\nu(U_1)$. This increases with $U_1$ from the value 
$\nu(U_1=1) = 1.3(1)$ to  $\nu(U_1=4) = 1.8(1)$.
We remark that the scaling ansatz \eqref{eqscal} is analogous to the one used for the 
standard MBL system with short-range interactions \cite{Luitz15}. The fact that the same scaling seems to hold
even in presence of long-range interactions suggests that  the underlying physics of our system is similar.

So far the gap ratio analysis together with the finite-size scaling 
indicates the existence of a boundary between {ergodic (ETH-like)} and MBL phase. The mean gap ratio reaches
the value characteristic for poissonian ensemble i.e. for integrable systems. This occurs for sufficiently large disorder values,
strongly dependent on the long range interaction strength $U_1$ which realizes  
  all-to-all couplings.

 For completeness, let us note that the considered  system possesses another non-ergodic phase
at large values of the long range interaction strength $U_1\gtrsim 10$ (not shown in Fig.~\ref{rstat}). 
This regime emerges when the all-to-all coupling term dominates in the Hamiltonian leading to non-ergodic dynamics
{due to global}
interactions
\cite{Lerose18}. We discuss this regime in Appendix B considering here the regime of small $U_1$.

\section{Dynamics of the system}
\label{dyna}

Imagine we prepare the system in a well defined {separable} state $|\psi_0 \rangle$. 
To probe the dynamics of the system we consider the time-dependent {density} correlation function 
\begin{equation}
 \label{eqcor1}
 C(t) = D \sum_{i=1}^K \left(  \bar n_i(t) - \bar n \right)\left( \bar n_i(0) - \bar n \right)\,,
\end{equation}
where $\bar n$ is the average number of particles and 
$\bar n_i(t)=\langle \psi(t) | n_i |\psi(t) \rangle$ is evaluated over the evolved state $|\psi(t) \rangle=\exp(-iHt)|\psi_0 \rangle$. Here, the constant $D$ warrants that $C(0) = 1$. According to  ETH, an ergodic system loses the memory of the initial state 
and the {correlation $C(t)$} decays to zero. Conversely, in the MBL 
phase the {density} correlation function $C(t)$ reaches a nonzero asymptotic value after a transient {time of the order of few $J^{-1}$ (which is here set to unity)}
\cite{Luitz15}.

The second quantity with which we probe the dynamics is the bipartite entanglement entropy $S(t)$. 
This is obtained after splitting the lattice into two subsystems A and B and calculating the density 
matrix {$\rho(t)$ of the subsystem $A$:
$\rho(t) = \mathrm{Tr}_B \{|\psi(t)\rangle\langle \psi(t)|\}$}, where $\mathrm{Tr}_B$ denotes the trace 
over subsystem B's degrees of freedom. The entanglement entropy is then defined as 
\begin{equation}
\label{ee1}
S=-\sum_i \rho_{ii}(t)\log(\rho_{ii}(t))\,, 
\end{equation}
where $\rho_{ii}$ are Schmidt basis coefficients squared with $\sum_i \rho_{ii}=1$ (see e.g. \cite{Karol}).
In systems with short-range interactions the logarithmic growth of the entanglement entropy $S(t)$ during {the} time evolution of the system
is a hallmark of MBL \cite{Bardarson12, Serbyn13a} and can be understood within the picture of LIOMs \cite{Serbyn13b, Huse14}. 
Systems with strong long-range interactions, on the other hand, manifest dynamical properties typical of ergodicity breaking, 
such as the logarithmic growth of the entanglement entropy after quenches even in absence of disorder
\cite{Daley2013,Daley2016,Fazio18,Lerose18}. In order to single out the onset of localization and 
the effect of long-range interactions on the localization properties, in the following subsections
we explore the transition {between non-ergodic and ergodic regimes} as a function of $U_1$ and for constant disorder 
amplitude $W=8$. We then turn to the regime where the long-range interactions are a weak perturbation to the dynamics. 

\subsection{Density correlations and entanglement entropy}
\label{correlations}

By inspecting the phase diagram (Fig.~\ref{rstat}), one can see that for 
the disorder strength $W=8$ the system is deep in the MBL phase for $U_1=0$. 
In fact, at $U_1= 0$ Eq. \eqref{H1} for fermions reduces to the XXZ Heisenberg spin chain and 
undergoes the ETH-MBL transition in the vicinity of $W_C(U_1=0)=3.7$ \cite{Luitz15}. This transition is 
accompanied by the appearance of {non-vanishing values of the correlation function at 
the asymptotics, $C(t\rightarrow \infty)\neq 0$,}  as well as {by the} logarithmic growth of the entanglement
entropy $S(t)$.
We now consider a nonvanishing value of the long-range interaction strength, and in 
particular analyse the cases (i) $U_1=1$, (ii) $U_1=3$, 
and (iii) $U_1=5$. The latter two cases are both in the localized regime for $K=16$, 
and yet they are delocalized in the thermodynamic limit according to the finite size scaling
in Fig.~\ref{rstat}. {The case $U_1=1$, instead, corresponds to a MBL phase for all system sizes we consider as well as in the thermodynamic limit (as predicted by the finite-size scaling, see Fig.~\ref{rstat}).} 

\begin{figure}[!ht]
\begin{center}
\includegraphics[width=1\columnwidth]{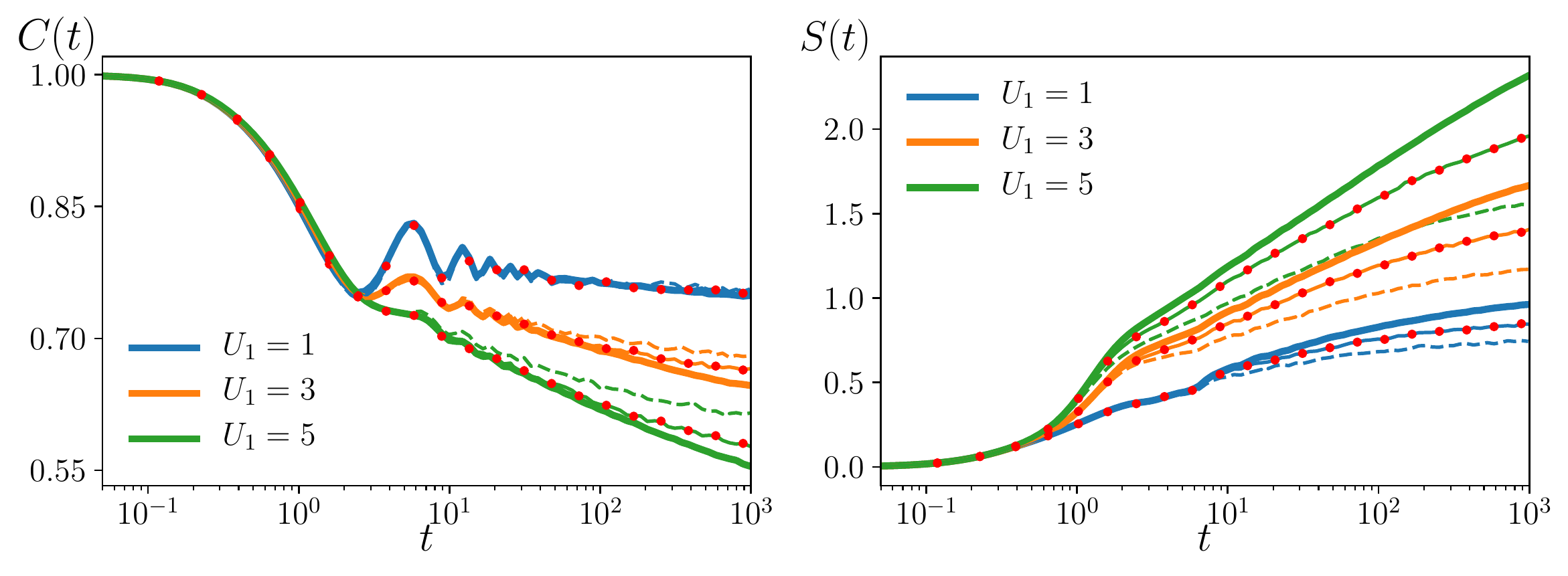}
\caption{ 
\label{cavdyn1}
Ergodicity breaking for the system of spinless fermions (half-filling) with 
lattice sizes $K=16,18,20$ (denoted respectively by dashed, dotted and thick lines), 
short-range interaction strength $U=1$, disorder strength $W=8$ and various 
long-range interaction strengths $U_1$. 
Left --  the correlation function $C(t)$,
Right -- the entanglement entropy $S(t)$. The quantities are averaged over
more than $2000$ disorder realizations, starting from randomly chosen initial Fock states $|\psi_0 \rangle$.}
\end{center}
\end{figure}

The time evolution of the correlation function $C(t)$ as well as of the entanglement entropy $S(t)$ 
is {displayed} in Fig.~\ref{cavdyn1}.  
We analyze system sizes $K=16,18,20$ using {the Chebyshev expansion technique for the time evolution}
(see Appendix \ref{cheby} for details).
For $U_1=1$ we observe the features characteristic of ergodicity breaking: 
 the correlation function $C(t)$ acquires a stationary value which very weakly depends on the system size
(compare {with the} left panel in Fig. \ref{cavdyn1}).
 The entanglement entropy $S(t)$ {(right panel}) shows an increase with time which is sublinear, and indeed seems weaker than logarithmic.
As the strength of long-range interactions $U_1$ increases  there appears a 
slow decay of the correlations $C(t)$ towards zero which becomes
more pronounced as the system size $K$ is increased.This result suggest that the correlations $C(t)$
decay to zero in the thermodynamic limit, which is in agreement with the 
results of finite-size scaling.
On the other hand the entanglement entropy $S(t)$ for {
$U_1=3$ and $U_1=5$} clearly grows logarithmically in time. Such a behavior is consistent with the picture of LIOMs and is believed to be a
feature of MBL system. {This seems to lead to an apparent paradox: In fact, while
the dynamics of $C(t)$ suggests} that large systems
would be ergodic, at the same time, the entanglement entropy growth $S(t)$ 
shows no signs of delocalized behaviour. Yet the behaviour of $S(t)$ could also
originate from the long-range nature of the interactions \cite{Lerose18}. We observe, in particular, 
that the slope of $S(t)$ increases with $U_1$ and with the system size. 
\begin{figure}[!ht]
\begin{center}
\includegraphics[width=1\columnwidth]{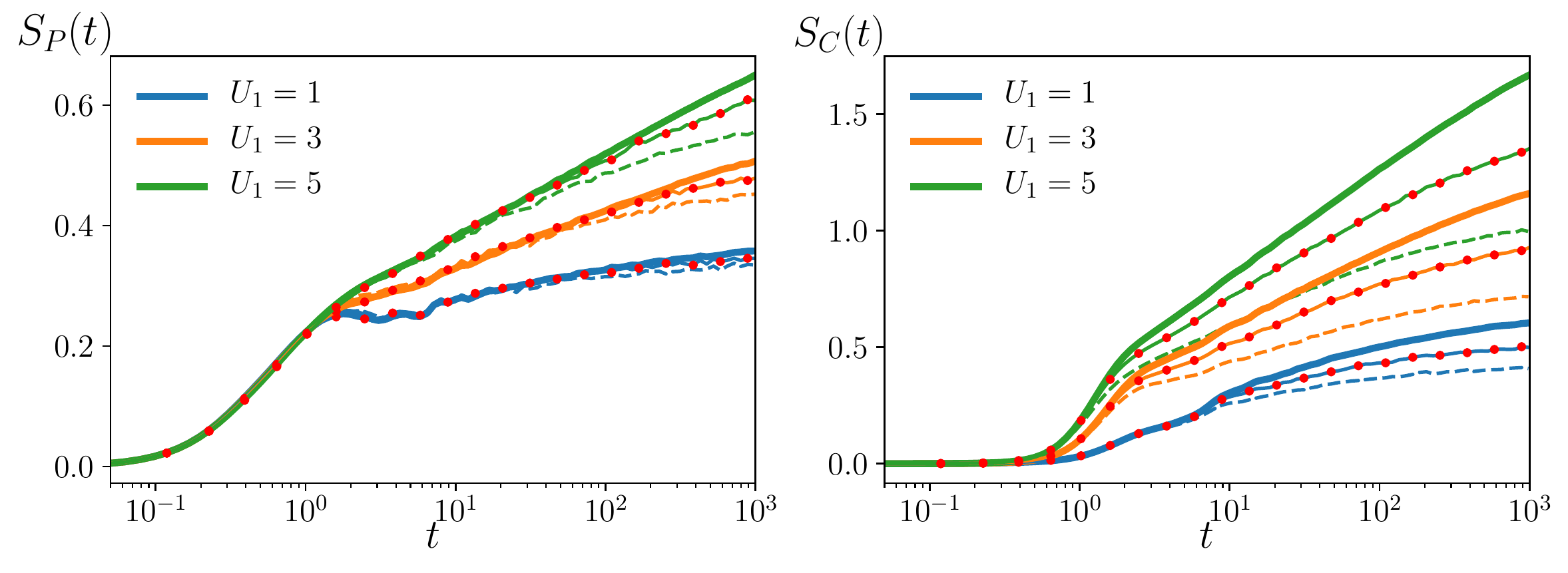}
\caption{
\label{cavdyn2}
Particle (left) and configuration (right) entanglement entropy ($S_P(t)$ and $S_C(t)$ respectively)
for the system of spinless fermions (half-filling) with 
lattice sizes $K=16,18,20$ (denoted respectively by dashed, dotted and thick lines), 
short-range interaction strength $U=1$, disorder strength $W=8$ and 
long-range interaction strengths $U_1=1,3,5$.}
\end{center}
\end{figure}

In order to gain insight we analyse in detail the behaviour of the entanglement entropy. 
Following Ref.~\cite{Lukin18} we express the entanglement entropy 
as the sum of two contributions $S(t)=S_P(t)+S_C(t)$, where $S_P(t)$ stems from particle 
number fluctuations between subsystems $A$ and $B$ and $S_C(t)$ is the entanglement 
entropy of different configurations of particles within the two subsystems. Let us denote 
by $p_n$  the probability of populating the $n$-particle sector in subsystem $A$ and by 
$\rho^{(n)}$ the corresponding block of the density matrix $\rho$ for subsystem $A$, such
that $\rho=\sum_np_n\rho^{(n)}$.
A simple manipulation shows that Eq. \eqref{ee1} can be rewritten as 
\begin{equation} 
S(t)=- \sum_{n=0}^N p_n \log(p_n) - \sum_{n=0}^N  p_n \sum_i \rho_{ii}^{(n)} 
\log(\rho_{ii}^{(n)}) \equiv S_P(t)+S_C(t).
\end{equation}
The resulting behaviors of $S_P(t)$ and $S_C(t)$ are shown in Fig.~\ref{cavdyn2}. 
We first notice that exchange of particles between subsystems $A$ and $B$ occurs due to tunneling. As visible in 
left panel of Fig.~\ref{cavdyn2} $S_P(t)$
grows significantly at a time scale of few tunneling times, independently of the value
of $U_1$ and of the system size. After this transient, its behaviour depends significantly on $U_1$
{and on the system size}. In particular, for 
$U_1=1$ it grows very slowly with time and weakly depends
on the system size, hinting towards a strong suppression 
of particle number fluctuations. For $U_1=5$, instead, it 
has a clear logarithmic growth in time  and a significant 
dependence on the system size. The former case is a standard MBL behavior \cite{Lukin18}:  the logarithmic growth of $S(t)$ is mainly due to 
the increase in the configuration entropy $S_C(t)$. The latter behavior, 
in which $S_P(t)$ grows logarithmically in time
enabling also faster and faster growth of $S_C(t)$ leads eventually to thermalization.
The dynamics at large $U_1$ thus points towards ergodicity for larger sizes, in agreement with the finite-size scaling analysis. Yet, it is so slow that  the decay time of the correlation function $C(t)$ is much slower
than in the ergodic regime at small $W$ and $U_1=0$.  We remark that  in standard models with short-range interactions 
deeply in the MBL phase, 
the time evolution can be efficiently simulated to large times ($\approx 10^3$) and 
for large systems sizes ($K\approx 10^3$) \cite{Zakrzewski18} with 
time-dependent density matrix renormalization group related approaches \cite{Schollwoeck11}.
{Such approaches are ruled out by the infinite interaction range of our model. 
We note that algorithms based on time dependent variation principle \cite{Haegeman11,Haegeman16} 
could in principle tackle large system sizes. We leave this task for future work and here consider small
sizes, amenable to diagonalization-like treatments.}

\subsection{Weak long-range interactions}
\label{weakLIOMs}

\begin{figure}[!ht]
\begin{center}
\includegraphics[width=1\columnwidth]{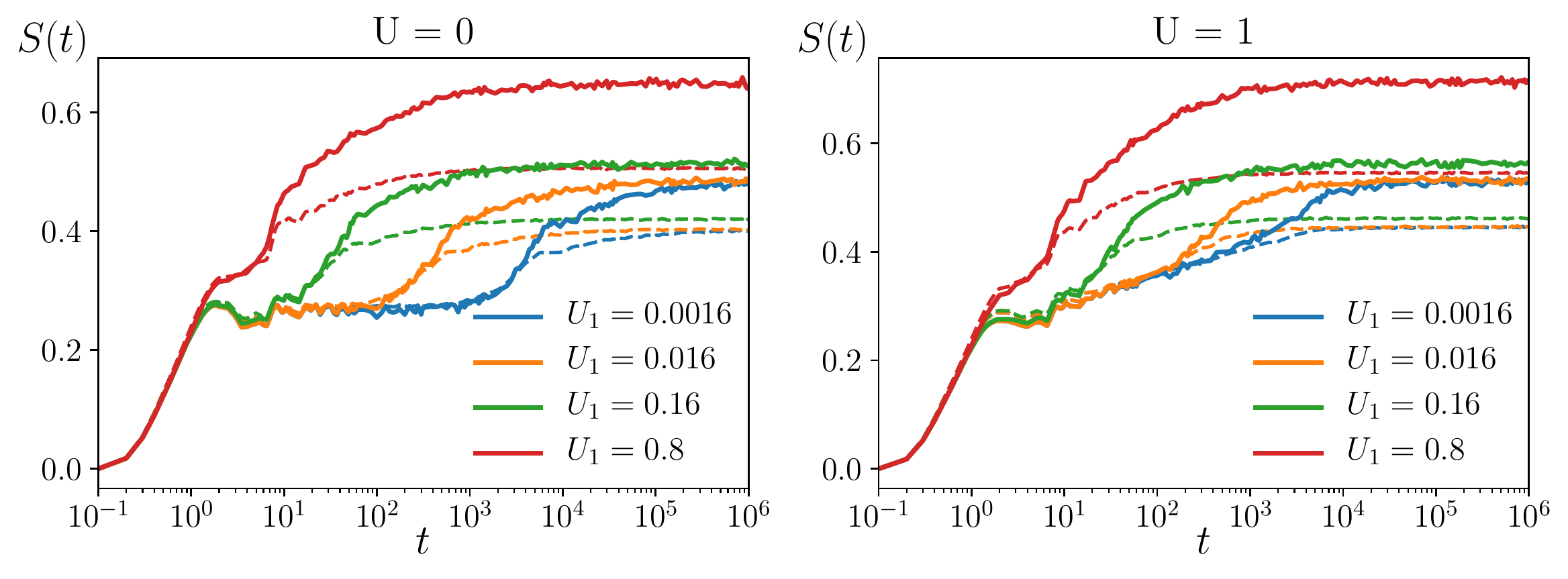}
\caption{\label{U1r0}
Entanglement entropy {as a function of time} for $U=0$ (left) and $U=1$ (right), $W=8$ and for different values of 
$U_1$ (see color code in the legenda). 
The dashed and thick lines correspond to $K=12$ and $K=16$, respectively.
}
\end{center}
\end{figure}

We now analyse the role of the long-range interactions {on MBL} by considering the limit in which $U_1$ is sufficiently weak with respect to the tunneling rate. We first focus on the case $U=0$, when the particles solely interact via the long-range interactions. For $U_1\ll 1$ we expect that the dynamics is first dominated by the hopping, and {only on a much longer time scale it is going to be visibly affected by $U_1$. 
The left panel of} Fig.~\ref{U1r0} displays the entanglement entropy $S(t)$ for different values of $U_1$, ranging from $1.6 \times 10^{-3}$ till $0.8$. The entanglement entropy first rapidly grows during an initial transient, which is {the same for all considered values of $U_1$ and is of the order of $1/J$ (which is here the unit of time)}. 
After this transient $S(t)$ saturates to a value for a time interval, up to a time scale 
$T_1\approx 1/U_1$. We understand this behavior  as the system being in the Anderson localization regime, 
since for this time scale the dynamics is the one of non-interacting particles. After $T_1$ the long-range
interactions start to significantly affect the dynamics and  the entanglement entropy $S(t)$ increases
approximately linearly with time, till it {saturates. The corresponding} saturation value depends 
on the given system size, on the strength of disorder $W$, and on the long-range interaction strength
$U_1$, {being, however, much lower, than the maximal (ergodic) value for a given system size.} 

The behaviour of the entanglement entropy $S(t)$ 
for $U=1$ (thus {when $U=J$}) is shown in the right panel of Fig.~\ref{U1r0}.
The time scale separation allows us to identify two behaviours characterizing the entanglement growth:
the growth which goes logarithmic in time, as for
a standard MBL system, and the rapid transient of $S(t)$ at the time scale $T_1$ due 
to the all-to-all coupling. The saturation values of the entanglement entropy seems to be 
only weakly dependent on
the interaction strength $U$. This leads us to conjecture that the origin of the nonergodic dynamics here is also MBL and it arises from 
a quasi-degenerate manifold of states coupled by the long-range interactions. 

\begin{figure}[!ht]
\begin{center}
\includegraphics[width=0.9\columnwidth]{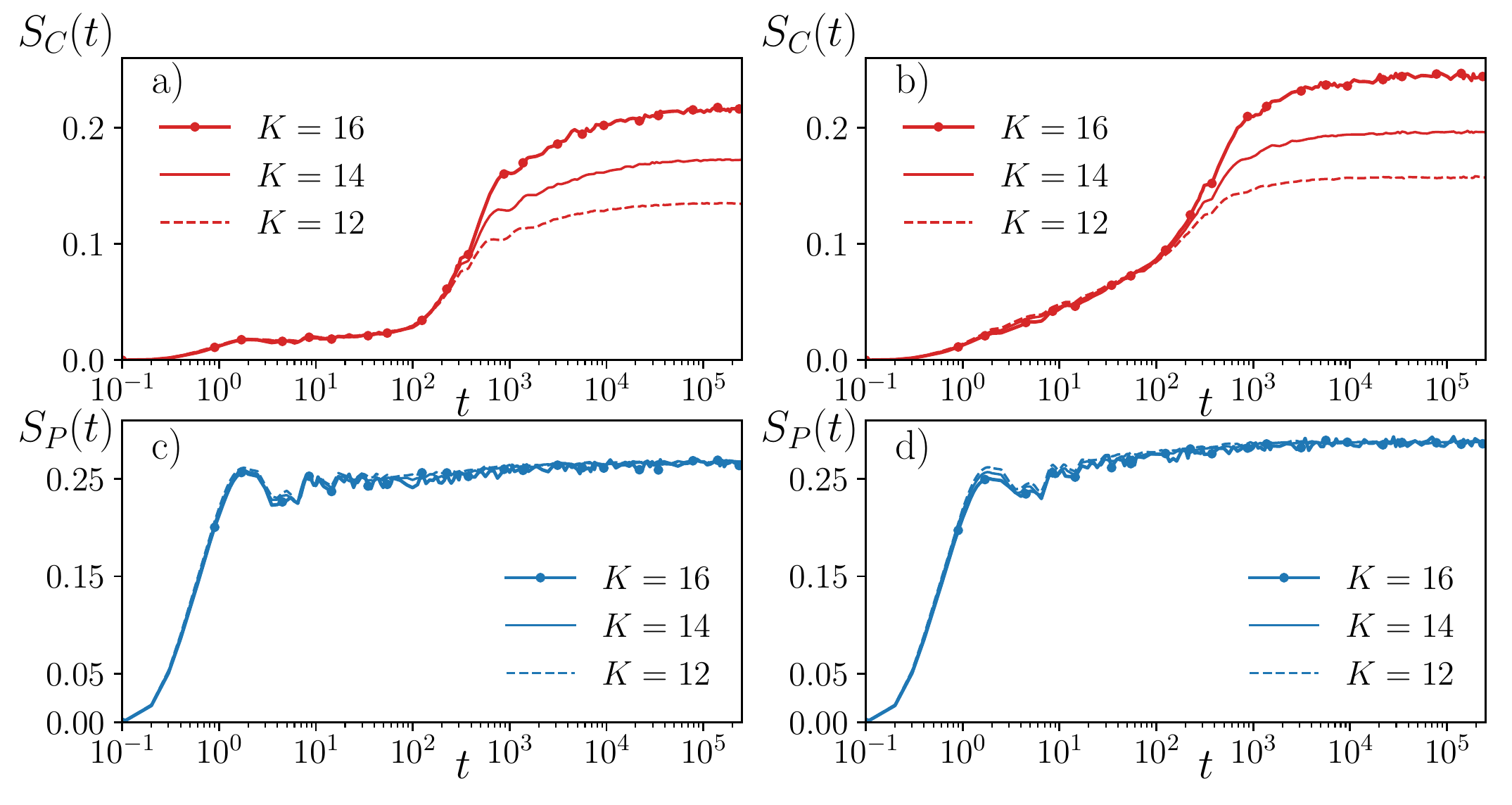}
\caption{\label{EE}
Configuration entanglement entropy $S_C(t)$ [top row, panels a) and b)] and the particle entropy $S_P(t)$ 
[bottom row, panels c) and d)] as a function of time  
in the localized regime ($W=10$) of the model with {$U=0$ and $U_1=0.016$ corresponding to the Anderson case perturbed by the long-range interactions}
[left column, panels a) and c)]. The plots in the right column [panels b) and d)] with $U=1$ 
correspond to long-range interaction perturbation of the MBL case. The effect of small $U_1$ in both cases is quite similar.
Thick, dashed and thin lines correspond to $K=16, 14$ and $K= 12$ respectively. 
}
\end{center}
\end{figure}

In order to test this conjecture, in Fig.~\ref{EE} we separately plot the particle and the configuration 
entanglement entropy as a function of time and corresponding to the curves {of Fig. \ref{U1r0} with $U_1=0.016$ and $U=0$ and $U=1$}.
For both values of $U$ the particle entanglement entropy $S_P(t)$ increases rapidly as the initial occupation
of {the} lattice sites spreads due to tunneling.  This transient dynamics occurs on a time scale of the order of $1/J$,
after which the particle number fluctuations change only marginally: observe that the associated $S_P(t)$ is 
approximately independent of the considered system sizes {$K$. Instead, the configuration entanglement entropy 
$S_C(t)$ shows different behaviors as a function of $U$ and of $K$. The dynamics of $S_C(t)$ can be characterized by the time scale $T_1\sim 1/U_1$: for $t\lesssim T_1$
and} for $U=0$ it is roughly constant, while for $U=1$ it grows logarithmically with time.
After $T_1$ {the configuration entanglement entropy} rapidly grows and then
saturates to a value which increases with the system size. We understand this increase as the number 
of accessible configurations grows with $K$. 
On the basis of this analysis we conclude that the dynamics for $U=0$ and small $U_1$ is a textbook case
of Anderson localization {perturbed by  weak long-range interactions. The long-range interactions couple only states that are
closely spaced in energy and that are localized in different regions of space.} 
Thus long-range interactions lead to a spread of the initial state
among relatively few localized eigenstates,
the corresponding dynamics is strongly nonergodic. 
The analogies shared between the dynamics at $U=0$ and $U=1$ for
small $U_1$ suggest that for $U\neq 0$ the dynamics is MBL-like. MBL is perturbed 
by the long-range interactions, which strongly couple the
quasi-degenerate manifold of localized states and at the same time the dynamics remains nonergodic. 

We explore the properties of this peculiar MBL phase by using a LIOM picture. We first recall that
the Hamiltonian of a generic (fully) many-body localized  system can be expressed as \cite{Serbyn13b, Huse14}
\begin{equation}
 \label{eqliom1}
 H  = \sum_{i=1}^K J^{(1)}_i \tau^z_i + \sum_{i,j=1}^K J^{(2)}_{ij} \tau^z_i \tau^z_j + 
 \sum_{i,j,k=1}^K J^{(3)}_{ijk} \tau^z_i \tau^z_j\tau^z_k + ...,
\end{equation}
where $\tau^z_i$ are quasi-local operators knows as LIOMs or l-bits. They can be thought of as dressed
occupation number operators as $\tau^z_i = U^{\dag} n_i U$ where $U$ is a quasi local unitary transformation.
The couplings $J_{ij}$ fall off exponentially with {the distance between interacting l-bits as }
\begin{equation}
\label{eqcoup1}
J^{(2)}_{ij} =J_0 \rm e^{-|i-j|/\xi}\,,
\end{equation}
{where $\xi$ is the localization length  (a similar relation} holds for higher order couplings
$J^{(3)}_{ijk}, ...$). It has been analytically shown \cite{Znidaric18}
that the l-bit model \eqref{eqliom1} leads to a logarithmic growth of the Renyi-2 entropy $S_2(t) = -\log \rm Tr \rho^2$
(which for large times and large system sizes {behaves essentially as the von-Neumann} entanglement
entropy $S(t)$) assuming that the initial state is an equal superposition of all Fock states. Moreover, to observe the 
logarithmic growth of the Renyi entropy $S_2(t)$ {it suffices to keep only the coupling coefficients $J^{(2)}_{ij}$, Eq. \eqref{eqcoup1},
setting the higher-order couplings $J^{(3)}_{ijk}, ...$ equal to zero}.

\begin{figure}[!ht]
\begin{center}
\includegraphics[width=0.6\columnwidth]{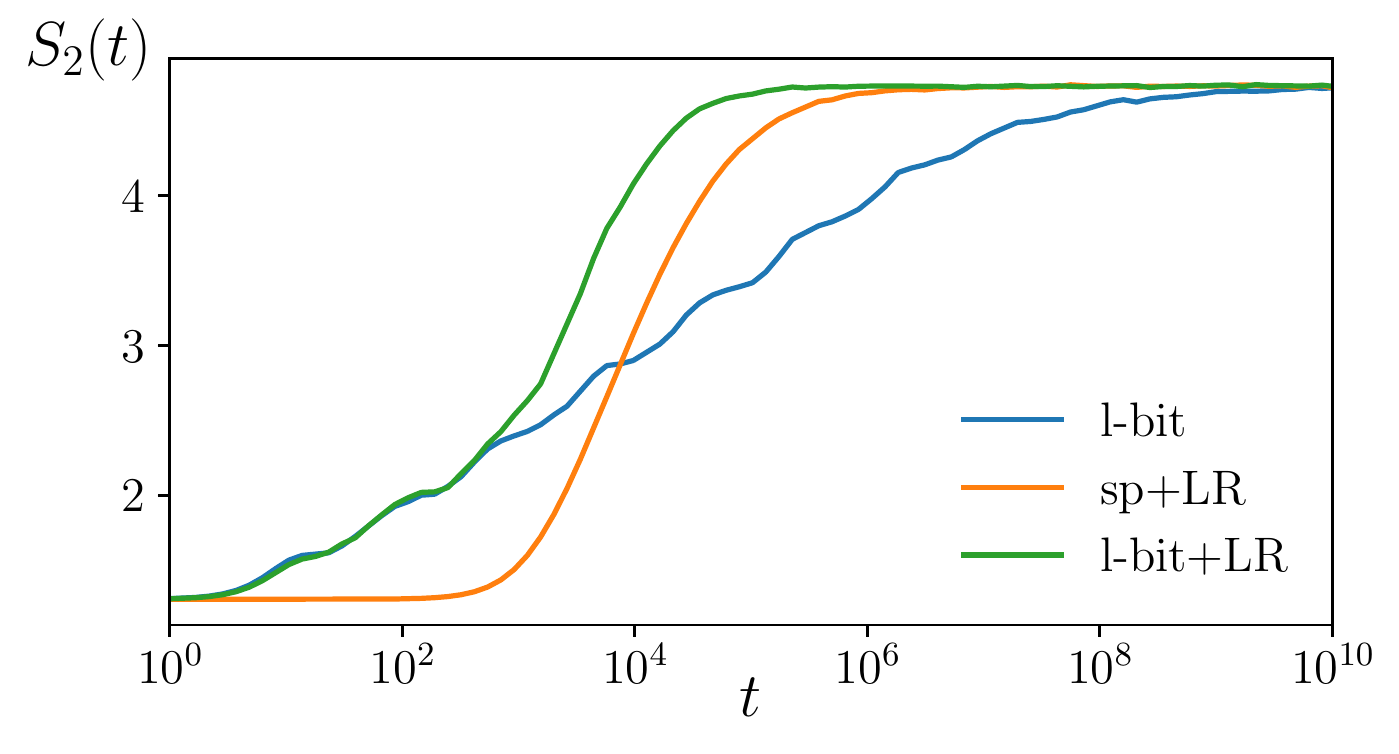}
\caption{\label{EEc} Renyi entropy $S_2(t)$ as a function of time for the $l-bit$ model of Eq. \eqref{eqliom1} with (i) $J_0 > 0$, $J_1 = 0$ (l-bit), (ii)
$J_0  >  0$, $J_1 > 0$ (l-bit+$LR$), (iii)  $J_0  =  0$, $J_1 > 0$ (sp+$LR$){, see legenda}.
The system size is $K=16$ and the initial state  (following \cite{Znidaric18}) is 
$|\psi\rangle=\sum_{i=1}^{N_H} |FS_i\rangle/\sqrt{N_H}$ where $|FS_i\rangle$ is $i$-th
state of the Fock basis and $N_H$ is dimension of the Hilbert space.
The value of Renyi entropy $S_2(t)$ corresponds in this model 
to $S_{C}(t)$ of the extended Hubbard model with spinless fermions.}
\end{center}
\end{figure}

In Fig.~\ref{EEc} we show that 
the Renyi entropy of the
l-bit Hamiltonian \eqref{eqliom1} reproduces the behaviour of the configuration entropy $S_C(t)$ for the
extended Hubbard model with spinless fermions. 
Specifically, assuming the exponential decay of the coupling terms $J^{(2)}_{ij}$, Eq.
\eqref{eqcoup1}, we reproduce the logarithmic growth of entanglement entropy characteristic of standard MBL. In order to
reproduce the rapid growth of $S_C(t)$, we introduce long-range couplings between LIOMs according to
\begin{equation}
\label{eqcoup2}
\tilde{J}^{(2)}_{ij} =J_0 \rm e^{-|i-j|/\xi} + \frac{J_1}{K} (-1)^{i+j} r_{ij}\,,
\end{equation}
where the term $J_1/K (-1)^{i+j} r_{ij}$ mimics the coupling experienced by l-bits
caused by the long-range interaction term $H_C$ \eqref{H_c} and $r_{ij} \in [0,1]$ is a random variable which models the overlap of $\tau^z_i$ and $n_i$. 
If we set $J_0 = J_1 =0$ in Eq.~\eqref{eqcoup2}, then there is no growth of entanglement entropy in the 
l-bit model which corresponds to no growth of configuration entropy -- a situation {characteristic} of Anderson localization.
Introducing non-zero $J_0$ in the model \eqref{eqliom1} we obtain the logarithmic growth of $S_2(t)$ -- the hallmark of MBL.
Setting then $J_1$ to a finite, non-vanishing value one gets a rapid growth of entanglement entropy which starts at a certain time scale
set by $J_1$: After this growth $S_2(t)$ saturates at the same value as in the case $J_1=0$.
All of those feature are  in qualitative agreement with the growth of configuration entropy $S_C(t)$
for our system at $U_1=0.016$ and $U=1$. 

The analysis of this Section strictly applies for very small values of $U_1$, such that the time scale $T_1 = 1/U_1$
is much larger than the time scale $J^{-1}$ set by the tunneling rate. 
As visible in Fig.~\ref{U1r0}, this separation of time scales takes place as long as $U_1 \lesssim 0.16$. The 
saturation value of the entanglement entropy $S(t)$ is already slightly larger for $U_1=0.16$ than for smaller values of $U_1$ 
meaning that a slight modification of structure of the LIOMs (possibly an increase of the support of $\tau^z_i$)
happened. For larger $U_1=0.8$, the saturation value of $S(t)$ is significantly larger. 
On the basis of the discussion so far, in particular of the studies in Sec. \ref{phasediagram}, we conclude that the system 
is still MBL, however, the properties of l-bits $\tau^z_i$ are significantly affected by the long-range interactions. 

{In summary, we have shown} that the cavity mediated long-range interactions lead to non-ergodic behavior 
of the system, which in presence of strong disorder may be interpreted as MBL. {However,}
the logarithmic growth of entanglement entropy, a hallmark of standard MBL is missing in our model, at least for 
very small $U_1$. The observed ergodicity breaking exhibits novel features, such as the rapid growth of entanglement entropy {which we attribute to} 
the long-range interactions. When the effect of the long-range interactions can be separated from 
the short-range interactions {in the dynamics}, then the entanglement entropy increases in good approximation linearly in time. 
Interestingly  the entanglement entropy is still bounded by a constant which only moderately changes
with $U_1$ as long as the system is in the localized phase. On the other hand, the dynamics and
in particular the growth of the configuration entropy may be explained using the language of LIOMs as
described above, the system is integrable (as the gap ratio analysis reveals poissonian statistics) 
and the entropy growth saturates {to smaller values than those found} for ergodic dynamics.
For these reasons we consider the behavior observed as a nonstandard variant of MBL. Let us also note that the observed phenomenon is  different from the 
non-ergodic phase observed for systems with single particle
mobility edge \cite{LiX15,LiX16,Hsu18,Kohlert18} where long-range interactions are not involved.

\section{Detection of ergodicity breaking: Light at the cavity-output}
\label{imbalance}

The dynamics of the system can be {experimentally revealed via a site-resolved measurement of $n_i$, which can be performed in  cold atoms experiments \cite{Schreiber15} and which allows one to reconstruct 
the correlation function $C(t)$}.  In this section we argue that the cavity setup can allow one to measure the breakdown of ergodicity {by photo-detection of the}
light at the cavity output.  {The emitted light, in fact, is scattered by the atoms and thus contains the information on their density distribution. In particular, the electric field amplitude, which we denote by $E_{\rm out}$, is directly proportional to the expectation value of population imbalance $I(t) =  \sum_i (-1)^{i} n_i$, Eq.~\eqref{Eq:I} in the dispersive
cavity regime (see Appendix A and \cite{Mottl11,Mottl12,Habibian13,Landig16,Hruby18}):}
\begin{equation}
E_{\rm out}(t)\propto \langle I(t)\rangle\,,
\end{equation}
which can be measured via heterodyne detection \cite{Mottl11} (Recall that {an observable similar to} $I(t)$ was 
employed in Ref. \cite{Schreiber15} to demonstrate MBL for a system of fermions in the optical disordered lattice).
{This measurement introduces projection noise that affects the atomic dynamics. Nevertheless, as we detail in Appendix A, this noise is negligible in the limit we consider, where the cavity field dynamics occurs on a much  faster time scale than the atomic motion}. The information extracted from the light at the cavity output has been used in this fashion to extract information about ground (or metastable) state phases of similar systems
\cite{Mottl12,Landig16,Hruby18}.
\begin{figure}[!ht]
\begin{center}
\includegraphics[width=.9\columnwidth]{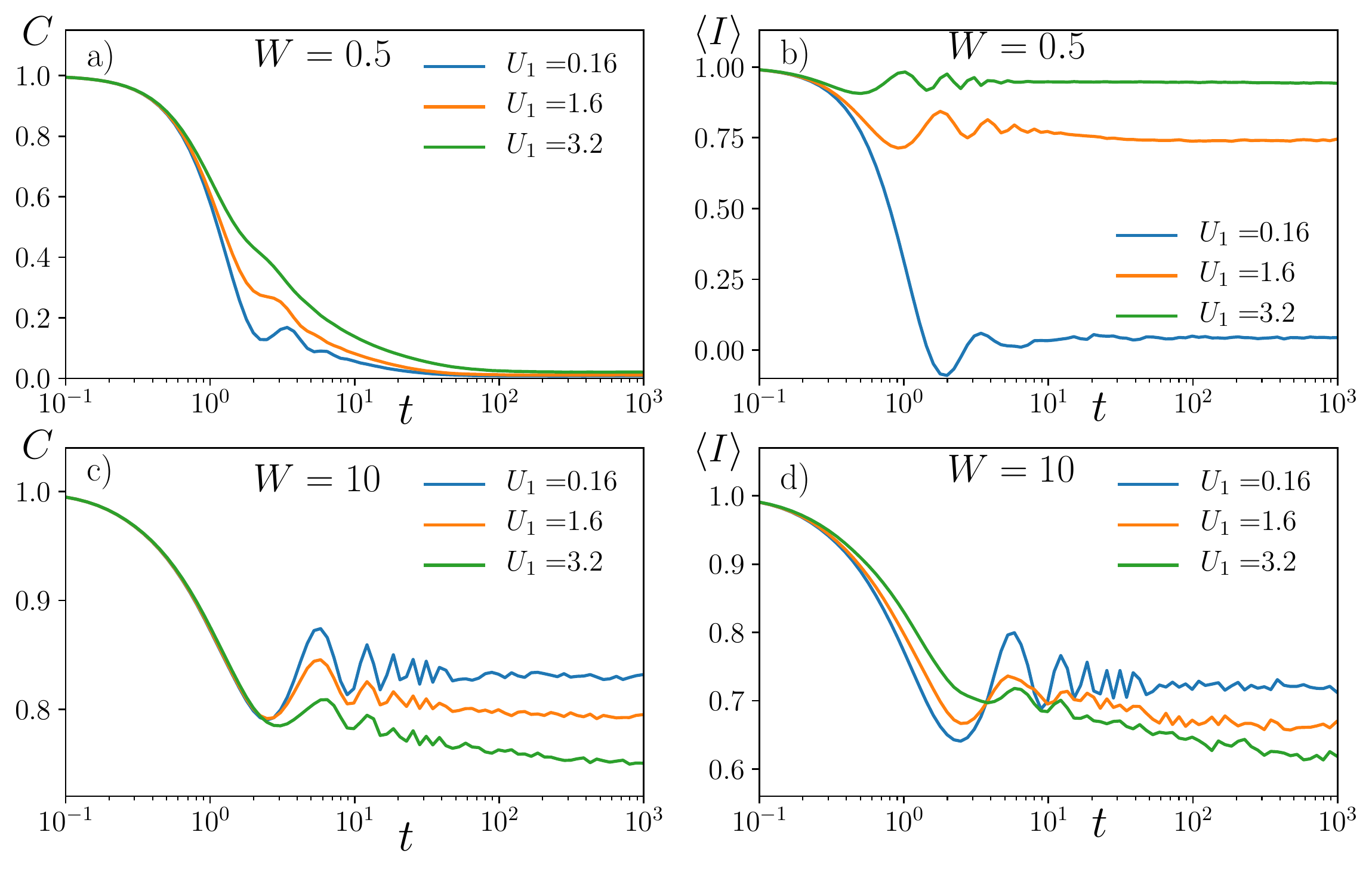}
\caption{Time evolution of the correlation function $C(t)$, Eq. (\ref{eqcor1}), 
[panels {a) and c)}] and the imbalance $\langle I\rangle$, Eq. \eqref{Eq:I} [
panels {b) and d)}] 
for different values of long-range interactions $U_1$. Top row [panels a) and b)]
corresponds to small disorder $W=0.5$ for which MBL is not expected. 
Yet, for  $U_1=1.6,3.2$ the imbalance saturates to finite values. 
The initial state of $C(t)$ is a random Fock state, 
for $\langle I \rangle$ is $|DW_{10}\rangle$. For strong disorder 
[bottom row, panels c) and d) ], where MBL is expected, 
the finite values of the population imbalance correspond to the
saturation of $C(t)$. The data have been calculated $K=16$  and averaged over $1000$
disorder realizations (starting time evolution from different randomly chosen Fock state 
in calculation of $C(t)$).}
\label{imbafig}
\end{center}
\end{figure}

Below we discuss the time evolution of the imbalance $\langle I \rangle$ and {of its square $\langle I^2\rangle$ (corresponding to the light intensity)
for two states} in which the system can be initially prepared. We first consider a density-wave like state $|DW_{10} \rangle = |101010..\rangle$, 
with odd sites occupied and even sites empty. This state maximizes {$\langle I \rangle$} for the fractional density $\bar n=1/2$ and minimizes
 the energies of both short-range and long-range interactions.
In the absence of disorder $|DW_{10} \rangle $
is an eigenstate of the Hamiltonian in the atomic limit \cite{Dogra16}.
{Starting from this state,} in the following we investigate to what extent the experimentally accessible 
time {evolution of the population imbalance $\langle I \rangle$}
allows one to probe ergodicity and its breaking in the system.
In this
analysis one shall keep in mind that {for  $U_1 \gtrsim W$ the state $|DW_{10} \rangle$ has a significant overlap with} the ground state, and its energy thus gets closer to the ground state {energy} as the ratio $U_1/W$ is 
increased. Figure~\ref{imbafig} diplays the time evolution of  $\langle I \rangle$ for 
the system initially prepared in  $|DW_{10} \rangle$ state. {The dynamics is shown for $W=0.5$ (upper panels) and $W=10$ (lower panels).
These two disorder strengths $W$} correspond to the ergodic and the MBL regimes, respectively, for 
the considered values of $U_1$. {For comparison, we also display the dynamics of the correlation function 
$C$ for the same parameters,} but when the initial state is a random Fock state with the same density. 
In the ergodic regime ($W=0.5$), the correlation function $C$ decays to zero signifying total
relaxation of the initial density profile regardless of the value of $U_1$. The dynamics of the
population imbalance for the initial state $|DW_{10} \rangle$ depends strongly on the ratio $U_1/W$.
For $W=0.5$ and $U_1=0.16$, the imbalance $\langle I \rangle$
decays to zero and is a valid probe of the ergodic 
properties of the system. For the same disorder amplitude and larger {values of $U_1$,} instead,  $\langle I \rangle$
saturates at a constant value, from which we infer that the state $|DW_{10}\rangle$ has already
significant overlap with the ground state of the system. In this regime, the time evolution 
of $\langle I \rangle$ is not a valid probe of ergodicity of a typical state of the system.
 
For strong disorder, $W=10$, on the other hand, the information about
localization properties of the system provided by the correlation function $C$ {is also visible in} the 
time evolution of the imbalance $\langle I(t) \rangle$, even for  
$U_1=3.2$. This behaviour is due to the fact that the energy spectrum 
of the system is much broader, such that even {for values of the interaction strength as large as
$U_1=3.2$} the 
state $|DW_{10}\rangle$ is still in a region of the spectrum with a relatively large density of states. 
 
\begin{figure}[!ht]
\begin{center}
\includegraphics[width=0.9\columnwidth]{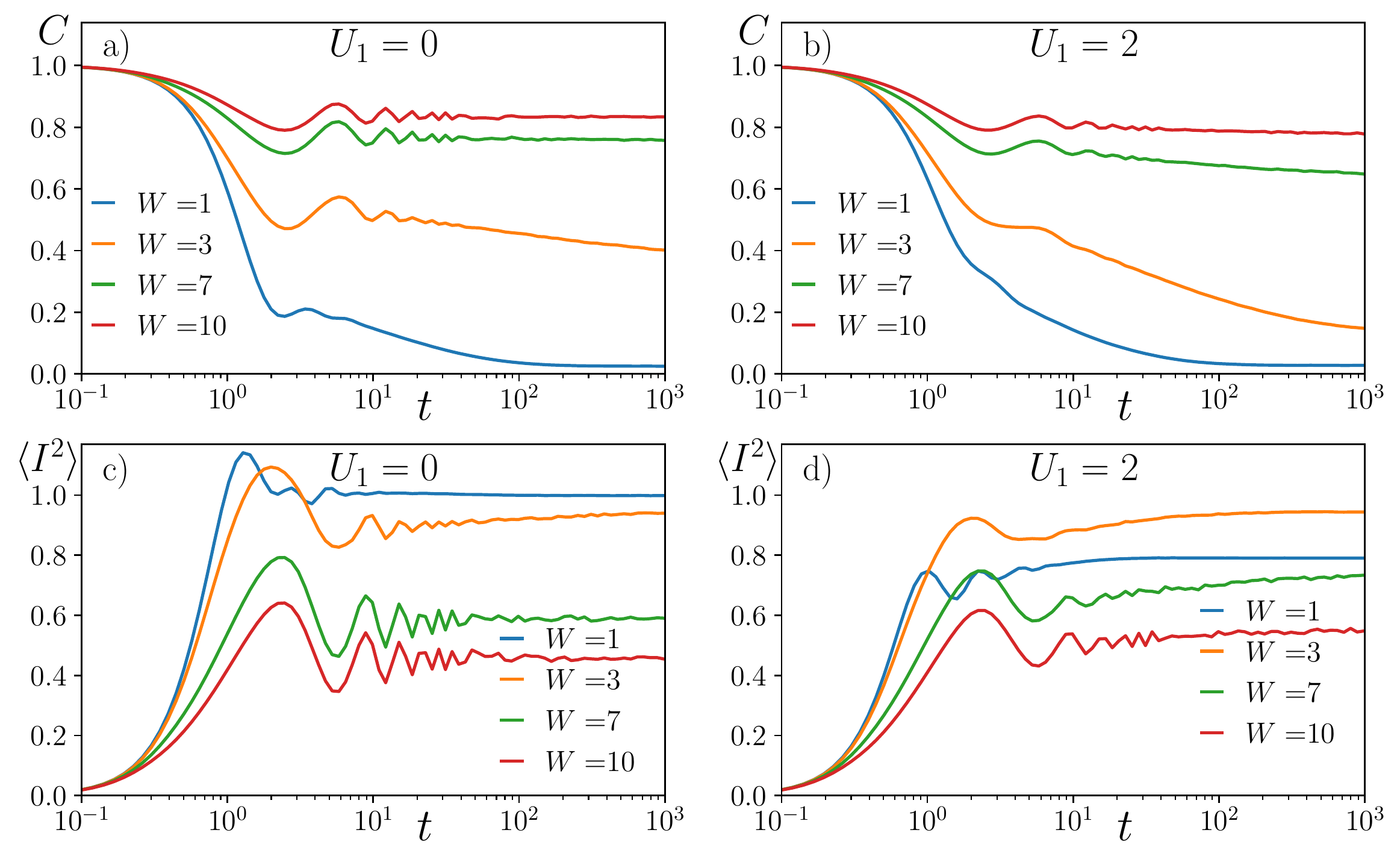}
\caption{{Time evolution of the correlation functions (top panels) and of $\langle I^2(t)\rangle$, 
corresponding to the intensity of the emitted light, (lower panels) 
for $U_1=0$ (a),(c) and $U_1=2$ (b),(d), and for different values of $W$. The initial state of the upper
panels is a random Fock state, for the lower panels state $|DW_{2}\rangle$. With increasing disorder strength the
correlations saturate to larger values in the MBL regime - that is associated with} a decrease of $\langle I^2\rangle$.
The normalization of $\langle I^2\rangle$ is chosen so that it saturates to unity at large times for $W=1$ and $U_1=0$.
The data have been calculated $K=16$ and averaged over $1000$
disorder realizations (starting time evolution from different randomly chosen Fock state 
in calculation of $C(t)$).}
\label{imbafig2}
\end{center}
\end{figure}

We now {consider the initial state} $|DW_2\rangle = |11001100...\rangle$, 
where the initial site occupations
form a pattern of singly-occupied and empty doublets of lattice sites. State $|DW_2\rangle$ 
is an eigenstate of the imbalance operator
$I$ {at the eigenvalue $0$, therefore its energy lies
about the band center}. Figure~\ref{imbafig2} {displays the time evolution of squared population imbalance $\langle I^2 \rangle$ when the initial state is 
$|DW_2\rangle = |11001100...\rangle$ and for different values of the disorder amplitude: $W=1,3,7,10$. The upper panels report, for comparison, 
the correlation function $C$ for the corresponding values of the disorder amplitude but when the initial state is a random Fock state.}
The dynamics of the density-density correlation function $C$ shows that the 
system is ergodic for $W=1,3$ with correlations eventually decaying to zero. {For $W=7,10$, instead, $C$ saturates to non-zero values at large times signalling MBL. The behaviours for vanishing long-range interactions and for $U_1=2$ are similar, see Fig.~\ref{imbafig2} a) and b).} If the system is ergodic, the ``intensity''
$\langle I^2 \rangle $ { increases from zero over a time scale of few $J^{-1}$ till it reaches a nonvanishing saturation value which depends on the disorder strength $W$. As visible in Fig.~\ref{imbafig2}
c), the saturation value is maximal in the ergodic phase at }$W=1$ and is smaller in the MBL phase
at $W=7,10$.
The saturation value of  $\langle I^2\rangle$ thus reflects the ergodicity  
of {the system dynamics. The saturation values of 
$\langle I^2 \rangle$ still point towards non-ergodicity for $U_1=2$ at $W=7,10$, consistently
with the value of mean gap ratio $\overline r$ for system size $K=16$, see Fig.~\ref{rstat}. Note that the
saturation values of $\langle I^2\rangle$ at $U_1=2$ are larger than
the ones reached for $U_1=0$ at the same disorder strength. This behaviour is consistent with the intuition that 
long-range interactions tend to delocalize. In Fig.~\ref{imbafig2}(d) one observes that the saturation value of $\langle I^2\rangle$ is lower for $W=1$
than for $W=3$. This suggests that the subspace of the
Hilbert space accessible during the time evolution from $|DW_2\rangle $ state is constrained by long-range interactions
at small disorder strength $W=1$, and that 
this constraint becomes weaker at $W=3$.}

{In summary, we have argued} that the measurement of light emitted by the cavity can 
allow one to determine the localization properties of the system by starting from well defined states. This probe of 
of ergodicity breaking is an appealing alternative to the band mapping technique 
of Refs. \cite{Sebby-Strabley06, Folling07} used in standard population imbalance measurements.

\section{Quasi-random disorder} 
\label{quasi}

Up till now we have considered a random on-site disorder.  Such a situation may be realized by an off-resonant pumping laser with a random intensity distribution.  This laser shall drive  an atomic transition
that does not scatter into the cavity field, and thus generate a random a.c. Stark shift of the onsite energy. 
The disorder may. also be  realized with the setups of Ref.~\cite{Landig16,Hemmerich} by introducing additional
weak laser beams creating optical lattices, whose periodicity is incommensurate with the cavity lattice periodicity. This setup creates quasi-random 
disorder analogously to the experiment of Ref. \cite{Schreiber15}. 
We now analyse localization in such a case i.e. when the onsite energy is due to the contribution of an incommensurate periodic 
potential, namely,
\begin{equation}
\label{eqquasi1}
 E_j = W \cos(2\pi \beta j + \phi)\,.
\end{equation}
Here, $W$ plays a role of disorder amplitude, $\beta$ is the ratio of the two lattice periodicities 
(we take $\beta = (\sqrt(5)-1)/2$) and the value of the phase $\phi \in [0,2\pi]$ 
determines the disorder realization. 
Quasiperiodic potentials such as the one of Eq.~\eqref{eqquasi1}
have been employed in a number of experiments investigating MBL in
optical lattices \cite{Schreiber15, Bordia16, Luschen17, Kohlert18}.
\begin{figure}[!ht]
\begin{center}
\includegraphics[width=1\columnwidth]{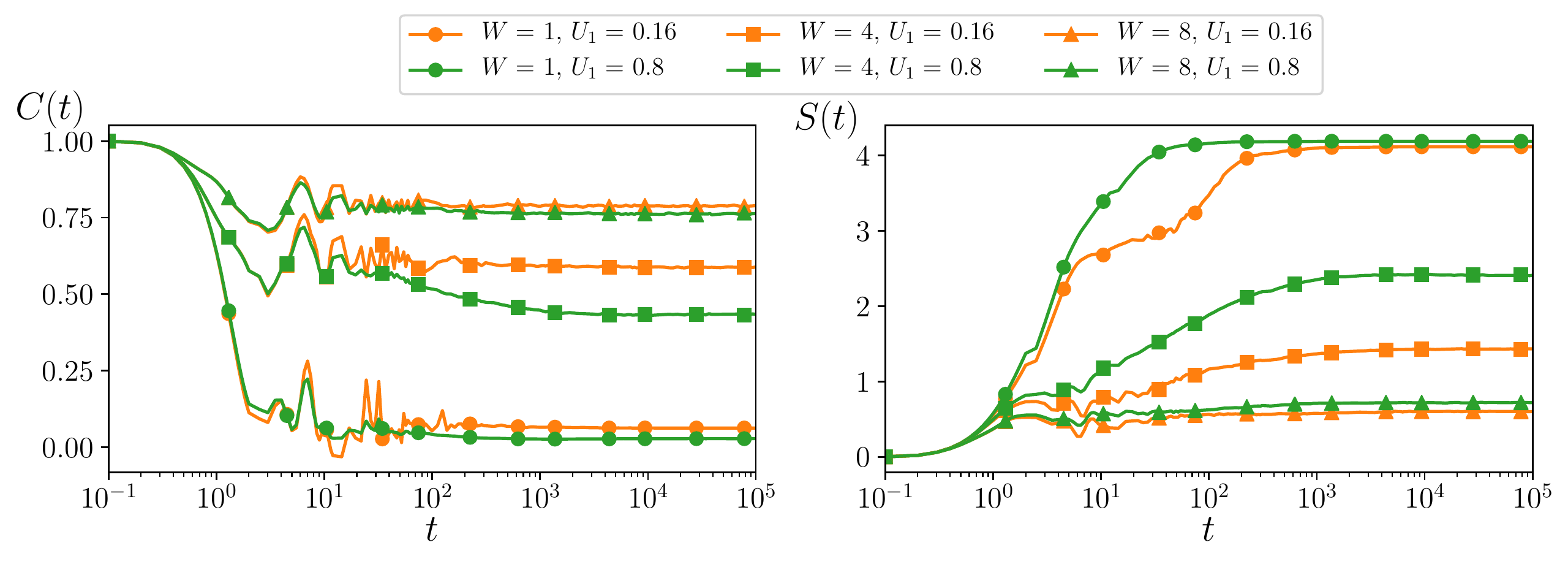}
\end{center}
\caption{\label{quasifig1} {Time evolution of the density correlation function $C(t)$ and of the} entanglement entropy 
$S(t)$ for $N=7$ spinless fermions on $K=14$ sites for the extended Hubbard model 
with quasirandom disorder, Eq. \eqref{eqquasi1}, for $U=0$ and the long-range interactions strengths
$U_1=0.16, 0.8$. The results are averaged over 1000 values of the phase $\phi$ in the interval $[0,2\pi]$, {the initial state is a Fock state with random site occupation and $N=7$}. 
 }
\end{figure}
\begin{figure}[!ht]
\begin{center}
\includegraphics[width=0.6\columnwidth]{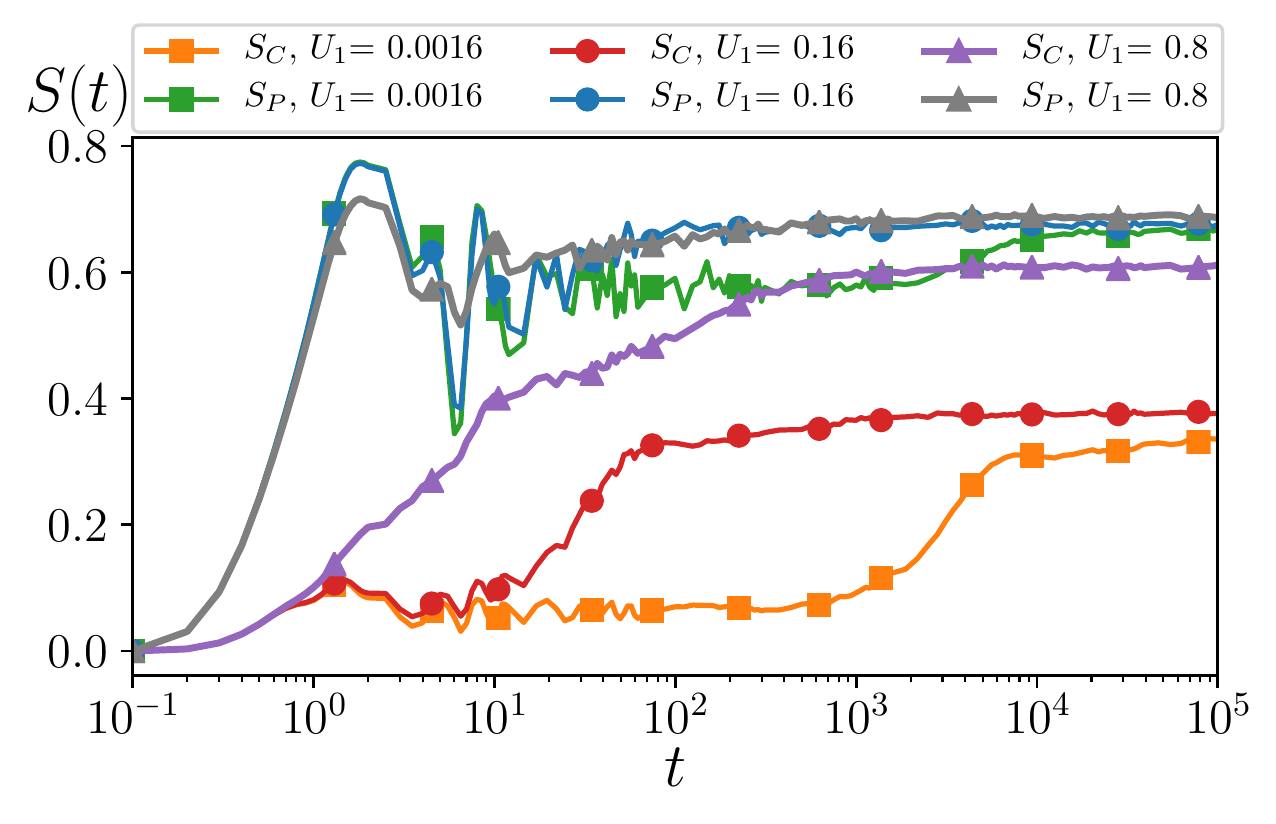}
\end{center}
\vspace{-1cm}
\caption{
\label{quasifig2}
{Time evolution of the configurational entanglement entropy $S_C(t)$ and of the particle entanglement entropy
$S_P(t)$ for $N=7$ spinless fermions} on $K=14$ sites with quasirandom disorder. The parameters are $U=0$ and $W=8$. 
The results {are} averaged over the phase $\phi$ in the interval $[0,2\pi]$, the initial state is a Fock state with random site occupation and $N=7$.
}
\end{figure}
It is well known that the properties of the MBL transition for purely random disorder
differ substantially from the ones of quasi-periodic disorder,
as it was shown by the analysis of the entanglement entropy \cite{Khemani17} 
and of the gap ratio \cite{Sierant18c}. Yet, while the important aspects 
of the transition itself are different, the ergodic and MBL phases are similar in the two settings, which is the reason
why the seminal observation of MBL \cite{Schreiber15} was feasible in a setup with quasiperiodic disorder. 
In order to show that {similar features are found} in our extended Hubbard model, we here discuss the dynamics of spinless fermions 
when the onsite energy in Eq. \eqref{H0} is given by Eq. \eqref{eqquasi1}.

The time evolution of the correlation function $C(t)$ and of the entanglement entropy $S(t)$ are {displayed} in Fig.~\ref{quasifig1}.
Two distinct regimes can be here identified: (i) the ergodic phase at $W=1$ in which the correlations $C(t)$ rapidly decays to zero
and entanglement spreads ballistically and (ii) the MBL phase at $W=8$ characterized by an asymptotic non-vanishing value of
the correlation function $C(t)$ as well as by the logarithmic growth of entanglement entropy. The intermediate disorder strength $W=4$ corresponds to the regime in which the localized-to-ergodic 
transition takes place as the strength of all-to-all coupling $U_1$ is increased.
To provide further evidence that the physics is similar for both random and quasiperiodic disorders, 
we calculate the time evolution of the configurational entanglement entropy $S_C(t)$ and of the particle entanglement entropy
$S_P(t)$. The results are presented in Fig.~\ref{quasifig2}. Similarly to the case of random disorder, small values of
$U_1$ lead to rapid growth of $S_C(t)$ at the time scale $T_1 =1/U_1$. For $U_1=0.16$, $S_C(t)$ reaches a larger asymptotic value. 
Further increase of the long-range interaction strength $U_1$ leads to the logarithmic growth of $S_C(t)$ with time, the entanglement entropy saturates at much larger value.

We thus predict that the properties of the MBL phase {of spinless 
fermions with quasiperiodic disorder and in the presence of cavity-mediated all-to-all interactions} are analogous to the ones {found when the disorder is instead random}.

\section{Bosons in the cavity}
\label{bosons}

\begin{figure}[!ht]
\begin{center}
\includegraphics[width=0.8\columnwidth]{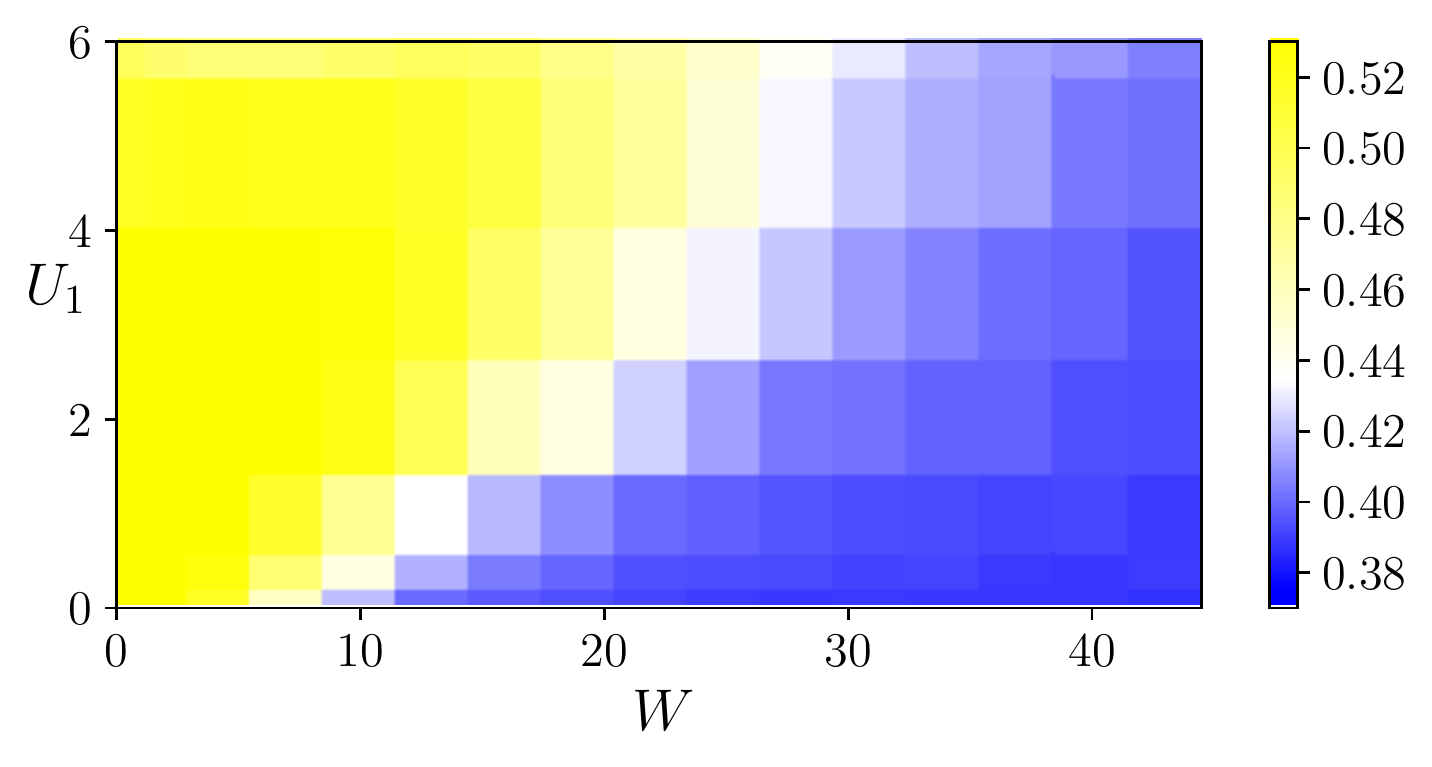}
\end{center}
\caption{\label{bosefig1}
{Contour plot of the average gap ratio $\overline r$ as a function of $W$ and $U_1$. The average gap ratio is calculated at the center of spectrum ($\epsilon = 0.5$)} for
$N=8$ bosons on $K=8$ lattice sites. The on-site interaction strength is
$U=1$, the results are averaged over 160 disorder realizations.}
\end{figure}

Many-body localization in bosonic systems was studied numerically 
in \cite{Sierant17, Sierant17b, Sierant18} as well as in experimental realizations
\cite{Lukin18, Rispoli18}. The essential features of the MBL phase in bosonic 
models are analogous to the system of spinless fermions (or spins). However, the 
possibility that the lattice site occupations exceeds unity leads to the natural appearance 
of many-body mobility edges at higher energy densities.
Motivated by the fact that {several experiments 
with cold atoms in optical resonators \cite{Landig16,Hemmerich} are performed with bosons, here we briefly discuss signatures of the MBL in}
the extended Bose-Hubbard Hamiltonian:
\begin{equation}
\label{H0b}
H=-J\sum_j^{K}\left(b_{j+1}^\dagger b_j+{\rm H.c.}\right)+\sum_j^K E_j n_j
+ U\sum_j^{K} n_j (n_{j}-1)-\frac{U_1}{K} \sum_{i,j}^{K} (-1)^{i+j} n_{i}n_{j}.
\end{equation}
Figure ~\ref{bosefig1} displays the phase diagram for a lattice consisting of $K=8$ sites at unit filling. The diagram {reports the mean gap ratio $\bar r$,
obtained from the gap ratio $r$ averaged 
over the states at the center of the spectrum $\epsilon \approx 0.5$. It shows} that for 
a given value of the all-to-all coupling $U_1$ there exists a disorder strength sufficient to localize the system.
The result is {similar} to the phase diagram of spinless fermions in Fig.~\ref{rstat}. We note however, that
the values of disorder {required to induce many-body localization in the bosonic system of Eq. \eqref{H0b} are larger
than the ones of} the fermionic counterpart. 
\begin{figure}[!ht]
\begin{center}
\includegraphics[width=0.99\columnwidth]{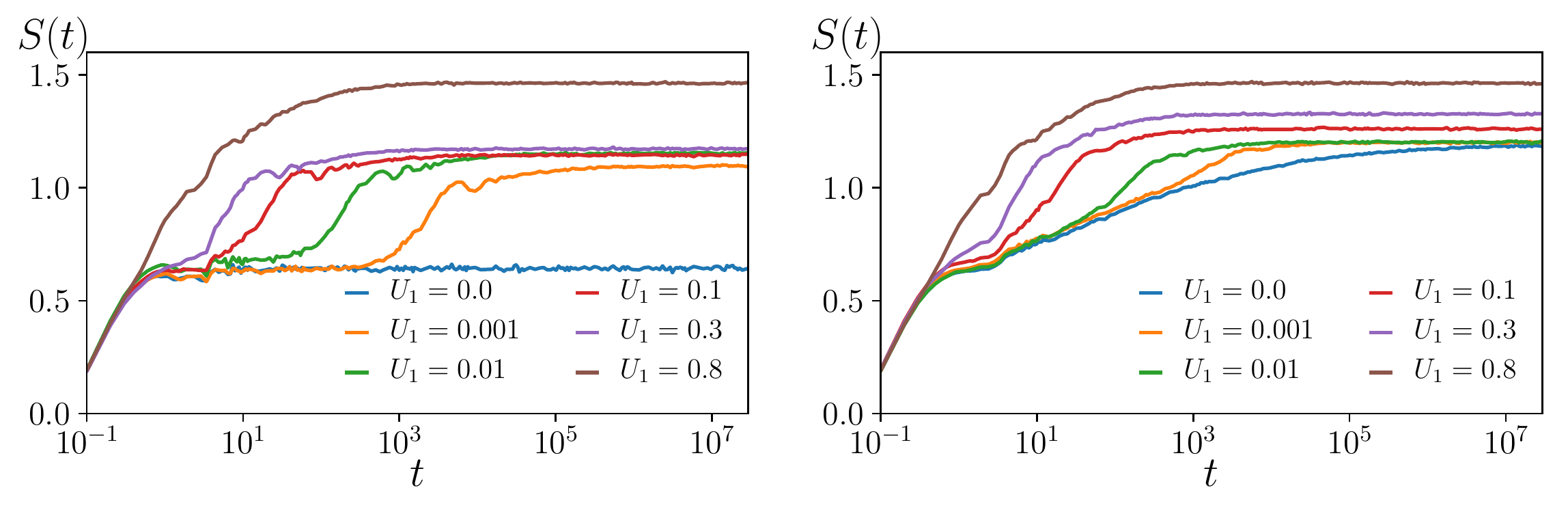}
\end{center}
\caption{\label{bosefig2} Bipartite entanglement entropy $S(t)$ {as a function of time} for
$N=8$ bosons on $K=8$ lattice sites. The left panel corresponds to $U=0$, the right panel to $U=1$, the initial
state is a random Fock state. The results are averaged over 2000 disorder realizations.}
\end{figure}

To provide further insight into {the physics of the MBL phase in all-connected} bosonic system
we calculate the bipartite entanglement entropy during the course of time evolution of the system -- c.f. 
Fig.~\ref{bosefig2}. The features are overall {similar to the ones found for} spinless fermions and thus seem to not be limited to small Hilbert spaces per site as for fermions.

\section{Conclusions}

In this work we have analysed the occurrence of many-body localization in a system of particles 
with all-to-all interactions. {The dynamics is described by an extended Hubbard model,
where the onsite energy follows a random distribution}. Our study is numerical and is based on exact
diagonalization as well as on methods for sparse Hamiltonian matrices.
By means of finite-size scaling we have shown that the MBL phase is indeed present  in
the system and {that the transition to the ergodic phase}
occurs at disorder amplitudes whose critical value increases with the
long-range interaction strength. 
The all-to-all interactions affect {the} spreading of the entanglement {during the time evolution: the growth of entanglement
entropy is faster than logarithmic in the MBL phase. It is instead linear when the tunneling coefficient is much larger than 
long-range interaction strength. Nevertheless,} the saturation
value of entanglement entropy is constrained by {particle number fluctuations between partitions} of the system.
 {A fixed distribution of particle number between subsystems is attained during the course of time evolution 
even in presence of all-to-all interactions and hence, the dynamics is non-ergodic.} 
We have shown that the features of the MBL phase can be qualitatively understood 
within {the} picture of LIOMs with long-range couplings.
The interplay between disorder and long-range interactions results in
transition from the MBL to an ergodic phase.
The characteristic features of the observed ETH-MBL transition seem to be independent of the 
quantum statistics, as the numerical analysis for finite systems of spinless fermions and of bosons show. 
{Similar} features are observed both for true random disorder as well as for quasi-periodic potential. 

This dynamics can be observed experimentally with quantum gases in an optical resonator. 
For these systems, indeed, {photo-detection of the light at the cavity output can allow one to probe the gas phase and thus} provide evidence of ergodicity breaking.

We finally remark that MBL phase in presence of global coupling to $d$ dimensional system
was found in \cite{Ng18}, particular scaling of $d$ with system size leads in high frequency limit to Hamiltonian 
similar to the one considered in this work.

\section*{Acknowledgements}
This work was performed with the support of EU via Horizon2020 FET project QUIC (nr.~641122). 
Numerical results were obtained with the help of PL-Grid Infrastructure. We acknowledge support of the
National Science Centre (PL) via project No.2015/19/B/ST2/01028 (K.B. and P.S.), 2018/28/T/ST2/00401 
(Etiuda scholarship -- P.S.) and the  QuantERA QTFLAG
programme No. 2017/25/Z/ST2/03029 (J.Z.). {G.M. acknowledges support by the German Research Foundation (DFG, Priority Program No. 1929 GiRyd)} and by 
the German Ministry of Education and Research 
(BMBF) via the QuantERA project NAQUAS is acknowledged. Projects NAQUAS and QTFLAG have received 
funding from the QuantERA ERA-NET Cofund in Quantum Technologies implemented within the European Union's Horizon 2020 Programme.

\begin{appendix}

\section{Derivation of the Hubbard Hamiltonian}
\label{Deriv-Ham}

Here we present the summary of the derivation of Hamiltonian (\eqref{H1}-\eqref{H_c}). The details in different 
variants of the model are discussed in e.g. Refs. \cite{Maschler08,Fernandez10,Mottl12,Habibian13b,Dogra16,Niederle16}.

We consider $N$ atoms of  mass $m$, they are confined within an optical cavity in a quasi-one-dimensional configuration colinear
with a one-dimensional optical lattice. The lattice oriented   along the cavity axis  has the same wave number $k=2\pi/\lambda$ as the cavity field. The atoms are prepared in the electronic ground state $|1\rangle$, the electric dipole transition $|1\rangle \to |2\rangle$ couples dispersively to the cavity and to the transverse laser with Rabi frequency $\Omega_z(x)$, see Fig. \ref{scheme}. Atoms and cavity field are treated in second quantization:  $\hat{a}$ and $\hat{a}^{\dagger}$ are the annihilation and creation operators of a cavity photon, respectively, and obey the commutation relation $[\hat{a},\hat{a}^{\dagger}]=1$; the atomic field operator $\hat\Psi_{j}(x,t)$ destroys an atom in the internal state $\ket{j=1,2}$ at position $x$ and time $t$, the commutation relations are $[\hat\Psi_i(x,t),\hat\Psi^\dag_j(x',t)]_{\pm}=\delta_{ij}\,\delta(x-x')$ (where $\pm$ indicate the anti- and commutation relations, depending on 
whether the atoms are fermions or bosons, respectively). Cavity and atomic operators commute at equal times. The Hamiltonian governing the dynamics can be decomposed into the sum of atoms, electromagnetic field, and atom-field interactions. The atomic Hamiltonian reads
\begin{eqnarray}
\hat{\mathcal H}_A
=\sum_{j=1,2}\int dx \,\hat\Psi^\dag_j(x)\hat{H}_j(x)\hat\Psi_j(x)+2U_{12}\int dx \,\hat\Psi^\dag_1(x)\hat\Psi^\dag_2(x)
\hat\Psi_2(x)\hat\Psi_1(x) \,,
\end{eqnarray}
where
\begin{equation}
\hat H_j(x)=-\frac{\hbar^2\partial_x^2}{2m}+V_{\rm cl}^{(j)}\cos^2(kx)+\frac{1}{2}\int dx'U_{jj}(x-x')\hat\Psi^\dag_j(x')\hat\Psi_j(x')-\hbar\Delta_a |2\rangle\langle 2|\,.
\end{equation}
Here, $\Delta_a=\omega_L-\omega_0$ is the detuning between the laser frequency $\omega_L$  and the  atomic transition frequency {in the frame rotating with the laser frequency $\omega_L$}, $V_{\rm cl}^{(j)}$ is the state-dependent depth of the optical lattice tightly binding atoms to its minima. The optical lattice potential is due to light shifts by a laser {driving a different atomic transition}. The other parameters are the collision rate $U_{jj}$, which is also state dependent, and the collision rate $U_{12}$ between atoms in different electronic states. 
Since the laser is here described by a classical field, the only quantum field is 
the cavity mode at frequency $\omega_C=ck$, with $c$ the speed of light. 
The field Hamiltonian in the frame rotating with the laser frequency takes the form
\begin{equation}
\label{H:c}
\hat{\mathcal H}_C=-\hbar\Delta_C\hat{a}^{\dagger}\hat{a}\,,
\end{equation}
where $\Delta_C=\omega_L-\omega_C$.
Finally, the Hamiltonian describing the interaction between the atomic dipoles and the electric fields is given by
\begin{eqnarray}
\label{Hint}
\hat{\mathcal H}_{\rm int}=\hbar \int dx \hat\Psi^\dag_2(x)\hat\Psi_1(x)(g(x)\hat a+\Omega_z(x))+{\rm H.c.}\,,
\end{eqnarray}
{where $g(x)=g_0\cos(kx)$ is the cavity vacuum Rabi frequency and $\Omega_z(x)$ is the position-dependent (real-valued) Rabi frequency of a laser propagating along the $z$ direction and 
coupling to the atomic transition (compare Fig.~\ref{scheme}).}

We now assume that the photon scattering processes are elastic, which is valid assuming that the largest frequency is the detuning $|\Delta_a|$. In particular, $|\Delta_a|\gg\gamma$, where $\gamma$ the radiative linewidth of the excited state, and $|\Delta_a|\gg \Omega,g_0\sqrt{n_{\rm cav}},|\Delta_C|$, namely, the detuning is much larger than the strength of the coupling between the ground and excited state, where $n_{\rm cav}=\langle \hat{a}^{\dagger}\hat{a}\rangle$ is the mean intracavity photon number. In this regime we can eliminate the excited state approximating  \cite{Larson08,Fernandez10}
\begin{align}\label{Psi_2}
{\hat\Psi}_2(x,t)\simeq&\frac{g(x)}{\Delta_a}\hat\Psi_1(x,t)\, \hat a(t)+\frac{\Omega_z(x)}{\Delta_a}\hat\Psi_1(x,t)\,,
\end{align}
which is valid to lowest order in the expansion in $1/|\Delta_a|$. Using Eq.~\eqref{Psi_2} in the Heisenberg equation of motion for the field operator $\hat\Psi_1(x,t)$ results in the equation (we now drop the subscript from the field operator: $\hat\Psi_1\equiv\hat\Psi$)
\begin{align}\label{Psi1_with_a}
\dot{\hat\Psi}=&\,-\frac{\rm i}{\hbar}[\hat\Psi,\hat{\mathcal H}_A]-{\rm i}\frac{\Omega_z(x)^2}{\Delta_a}\hat\Psi -{\rm i}U_0(x)\ \hat a^\dag\hat\Psi \hat a  -{\rm i}S(x) \Big(\hat a^\dag\hat\Psi+\hat\Psi \hat a\Big)\,,
\end{align}
where we have taken care to keep the ordering between cavity and atomic operators. This equation is now solely coupled to the Heisenberg-Langevin equations for the cavity field:
\begin{align}\label{a_with_Psi1}
\dot{\hat a}=&-\kappa \hat a+{\rm i}\left(\Delta_C-\int dx U_0(x)\hat n(x)\right) \hat a-{\rm i}\int dx S(x)\hat n(x)+\sqrt{2\kappa}\hat{a}_{\rm in}\,,
\end{align}
which depends on the atomic operators through the atomic density
 \begin{equation}
 \hat n(x)=\hat\Psi^\dag(x)\hat\Psi(x)\,.
 \end{equation}
Equation \eqref{a_with_Psi1} includes the quantum noise due to the cavity losses, with $\kappa$ the cavity loss rate and $\hat{a}_{\rm in}(t)$ is the input noise operator, with $\langle \hat{a}_{\rm in}(t)\rangle=0$ and $\langle \hat{a}_{\rm in}(t)\hat{a}_{\rm in}^\dagger(t')\rangle=\delta(t-t')$ \cite{Carmichael}. The other parameters are the frequency $U_0(x)=g(x)^2/\Delta_a$, which scales the depth of the intracavity potential generated by a single photon, and the scattering amplitude $S(x)=g(x)\Omega_z(x)/\Delta_a$ \cite{Maschler08,Larson08}.

We now  eliminate the cavity degrees of freedom from the atomic dynamics by identifying the time-scale $\Delta t$ over which the atomic motion does not significantly evolve while the cavity field has relaxed to a state which depends on the atomic density at the given interval of time. This is verified when $\Delta t\gg T_c$, where $T_c=1/|\Delta_C+{\rm i}\kappa|$, and $\sqrt{\omega_R E_{\rm kin}}\ll \hbar/\Delta t$, with $\omega_R=\hbar k^2/(2m)$ the recoil frequency and $ E_{\rm kin}$ the mean kinetic energy \cite{Rojan16}. We also require that the coupling strengths between atoms and fields, which determine the time scale of the evolution due to the mechanical effects of the interaction with the light, are much smaller than $1/\Delta t$. In this limit, we replace the field operator with its coarse-grained average $\hat{a}_{\rm st}(\bar t)$, defined as
$$\int_{\bar t}^{\bar t+\Delta t}\hat{a}(\tau)d\tau/\Delta t= \hat{a}_{\rm st}(\bar t)\,,$$
such that  $\int_{\bar t}^{\bar t+\Delta t}\dot{\hat a}_{\rm st}(\tau)d\tau=0$, with $\dot{\hat a}$ given in Eq. \eqref{a_with_Psi1}. The "stationary" cavity field is a function of the atomic operators at the same (coarse-grained) time, and in particular it takes the form
\begin{equation}\label{a_mf}
\hat{a}_{\rm st}(\bar t)=\frac{\int dx S(x)\hat n(x,\bar t)}{(\Delta_C-\int dx U_0(x)\hat n(x,\bar t))+{\rm i}\kappa}+\frac{{\rm i}\sqrt{2\kappa}\hat{\bar a}_{\rm in}(\bar t)}{(\Delta_C-\int dx U_0(x)\hat n(x,\bar t))+{\rm i}\kappa}\,,
\end{equation}
with $\hat{\bar a}_{\rm in}$ the input noise averaged over $\Delta t$, such that in the coarse-grained time scale $\langle \hat{\bar a}_{\rm in}(\bar t)\rangle=0$ and $\langle \hat{\bar a}_{\rm in}(\bar t)\hat{\bar a}_{\rm in}^\dag(\bar t')\rangle=\delta(\bar t-\bar t')$. Note that the commutation relations of the new operators are modified, and in particular the commutator between $\hat{a}_{\rm st}$ and $\hat{a}_{\rm st}^\dag$ scales as $T_c/\Delta t$, The quantum noise term is an effective term which provides the same averages as the corresponding quantum-noise operator one gets by formally integrating Eq. \eqref{a_with_Psi1}. It can be neglected when the mean intracavity photon number is larger than its fluctuations or by integrating for sufficiently long time. In this limit the dynamics is coherent and the field at the cavity output, 
\begin{equation}
\label{a:out}
\hat{\bar a}_{\rm out}(t)=\sqrt{2\kappa}\hat{a}_{\rm st}-\hat{\bar a}_{\rm in}\,,
\end{equation}
allows one to monitoring the state of the atoms \cite{Carmichael,Mekhov07,Rist10,Habibian13b}. 

Using Eq.~\eqref{a_mf} in place of the field $\hat{a}$ in Eq. \eqref{Psi1_with_a}, and discarding the noise term, leads to an equation of motion for the atomic field operator which depends solely on the atomic variables\cite{Larson08,Fernandez10}:
\begin{align}\label{Psi11_with_a}
\dot{\hat\Psi}=&\,-\frac{\rm i}{\hbar}[\hat\Psi,\hat{\mathcal H}_A]-{\rm i}\frac{\Omega_z(x)^2}{\Delta_a}\hat\Psi-{\rm i}\frac{\int dx' S(x')\hat n(x')}{\Delta_C}S(x)\hat\Psi(x)-{\rm i}S(x)\hat\Psi(x) \frac{\int dx' S(x')\hat n(x')}{\Delta_C}\,,
\end{align}
where we have assumed $|\Delta_C|\gg |U_0|n,\kappa$ and thus discarded the terms {scaling with $U_0$ in the denominators of Eq.~\eqref{a_mf}} (see \cite{Fernandez10} for a discussion). The {term of the right-hand side of Eq. \eqref{Psi11_with_a}} can be cast in terms of the commutator between $\hat\Psi$ and the effective Hamiltonian $H=H_A+H_{\rm CQED}$, where\,\cite{Fernandez10}:
\begin{equation}
H_{\rm CQED}=\hbar\int dx\,\frac{\Omega_z(x)^2}{\Delta_a}\hat n(x)+\frac{1}{\Delta_C}\hbar\left(\int dx\,S(x)\hat n(x)\right)^2\,,
\end{equation}
{which contains an infinitely ranged density-density interaction. This interaction} is attractive for $\Delta_C<0$, which is the regime we assume in this work.

The Hubbard model is obtained when the external optical lattice is sufficiently deep to tightly bind the atoms in the lowest band and when the cavity interactions are a sufficiently small perturbation. We denote by $K$ the number of lattice sites and perform the Wannier decomposition of the atomic field operator $ \hat\Psi(x)=\sum_{i}w_{i}(x) \hat b_{i}$ \cite{Jaksch98,Bloch08}. {Here, $w_{i}(x)$ denotes the  Wannier function of the classical optical lattice that is centered at  lattice sites  $x_i=ia$, with $a=\lambda_0/2$ the lattice periodicity}, while $\hat b_i$ annihilates a particle at the corresponding lattice site. We further assume the scaling $g(x)=\tilde g(x)/\sqrt{K}$ (thus $S(x)=\tilde S(x)/\sqrt{K}$), which is equivalent to assuming that the cavity mode volume scales linearly with the size of the lattice \cite{Brandes03,Fernandez10}. This scaling is equivalent to Kac's scaling and warrants that the energy is extensive despite the all-to-all interactions (even though it remains non-additive). The 
Hubbard term due to the cavity global interactions takes the form (recall $\Delta_C<0$) \cite{Habibian13b,Dogra16}:
\begin{eqnarray}
-\frac{1}{|\Delta_C|}\left(\int dx\,S(x)\hat n(x)\right)^2\approx -\frac{1}{K|\Delta_C|}\sum_{i,j}\left(\int dx \tilde{S}(x)w_i(x)^2\right)\left(\int dx \tilde{S}(x)w_j(x)^2\right)\hat n_i \hat n_j\,,
\end{eqnarray}
with $\hat n_i=\hat b_i^\dag\hat b_i$ {the occupation of site $i$}. For $\tilde{S}(x)=S_0\cos(kx)$, corresponding to a homogeneous transverse pump $\Omega_z(x)=\Omega$, {then} $S_0=\tilde{g}_0\Omega/\Delta_a$ and
\begin{equation} 
\int dx \tilde{S}(x)w_i(x)^2=S_0\int dx \cos(kx)w_i(x)^2=(-1)^iS_0\left|\int dx \cos(kx)w_i(x)^2\right|\,.
\end{equation}
Thus, one obtains Eq.\,\eqref{H_c} with the coefficient
\begin{equation}
U_{\rm l}=\frac{S_0^2}{|\Delta_c|}\left(\int dx \cos(kx)w_i^2\right)^2\,
\end{equation}
while the remaining terms in the tight binding approximation yield the Hubbard model. Let us note that the (coarse-grained)
field $\hat{\bar a}_{\rm out}(t)$ may be expressed {as (neglecting the noise term in the coarse-grained limit)}
\begin{equation}
\label{a:out2}
\hat{\bar a}_{\rm out}(t)=\frac{\sqrt{2\kappa}S_0}{\Delta_c\sqrt{K}}\sum_i(-1)^i\hat n_i-\hat{\bar a}_{\rm in}\,.
\end{equation}
{Hence the mean electric field at the cavity output is proportional to the expectation value of the population imbalance,
\begin{equation}
E_{\rm out} (t) \propto  \langle \sum_i (-1)^i \hat n_i\rangle= \langle I \rangle(t)
\label{a:out3}
\end{equation}
where we used the definition in Eq.~\eqref{Eq:I}. We remark that, for the same reason that the time-scale separation allows one to derive a Hamiltonian dynamics for the atomic motion, thus discarding the commutation relation between the cavity field operators, then projection noise due to cavity leakage and photo-detection can be neglected because it is of the same order as the non-adiabatic corrections.}

\section{Nonergodic regime for strong all-to-all interactions}
\label{nonergo}

\begin{figure}[!ht]
\begin{center}
\includegraphics[width=0.8\columnwidth]{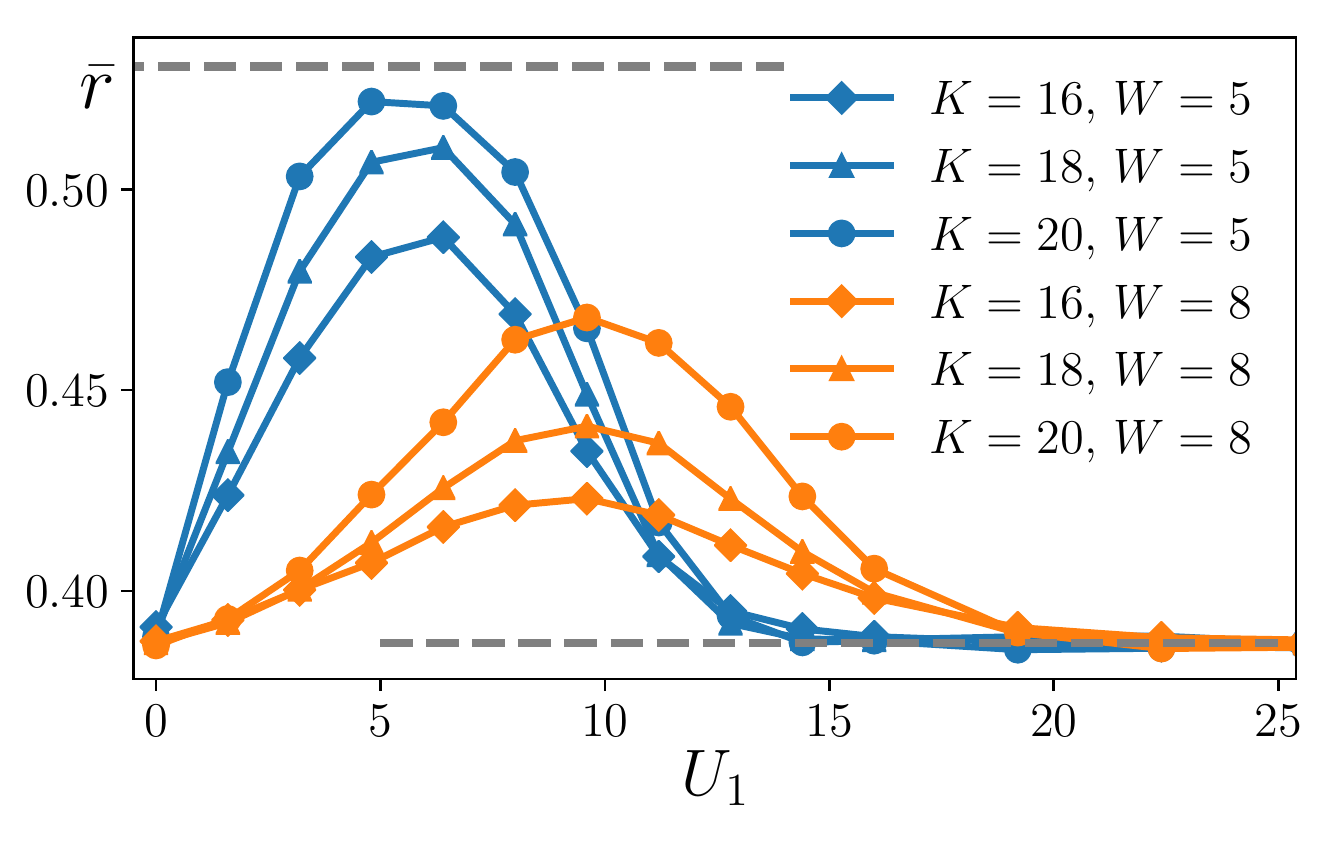}
\end{center}
\caption{\label{nonerg1} Mean gap ratio $\overline r$ {as a function of}
all-to-all coupling $U_1$ for different 
system sizes {and fixed disorder strength $W$}. 
Observe that for a given  {disorder strength $W$},
{the mean gap ratio $\overline r$ first increases, consistently with an appearance of the ergodic
phase shown in Fig.~\ref{rstat}. Then, for large $U_1$, the gap ratio tends back towards the value 
corresponding to Poisson statistics indicating an occurrence of another non-ergodic phase. 
The mean gap ratio is determined in 
the center of the spectrum 
(for $\epsilon_n = (E_n - E_{\rm min})/(E_{\rm max} -E_{\rm min})\approx 0.5$).} }
\end{figure}
Even in the absence of disorder strong long-range interactions may lead to nonergodic
behavior manifesting itself e.g. in the logarithmic or sublinear growth of the entanglement 
entropy for a sudden quench \cite{Schachenmayer13,Buyskikh16,Lerose18}. This behavior seems 
to be distinct from MBL. An appearance of such a regime in the model studied here may be 
visualized considering mean gap ratio for large $U_1$ values as shown in Fig.~\ref{nonerg1}.
For large $U_1$ system tends towards gap ratio $\overline r$ corresponding to Poisson statistics.

\begin{figure}[!ht]
\begin{center}
\includegraphics[width=0.99\columnwidth]{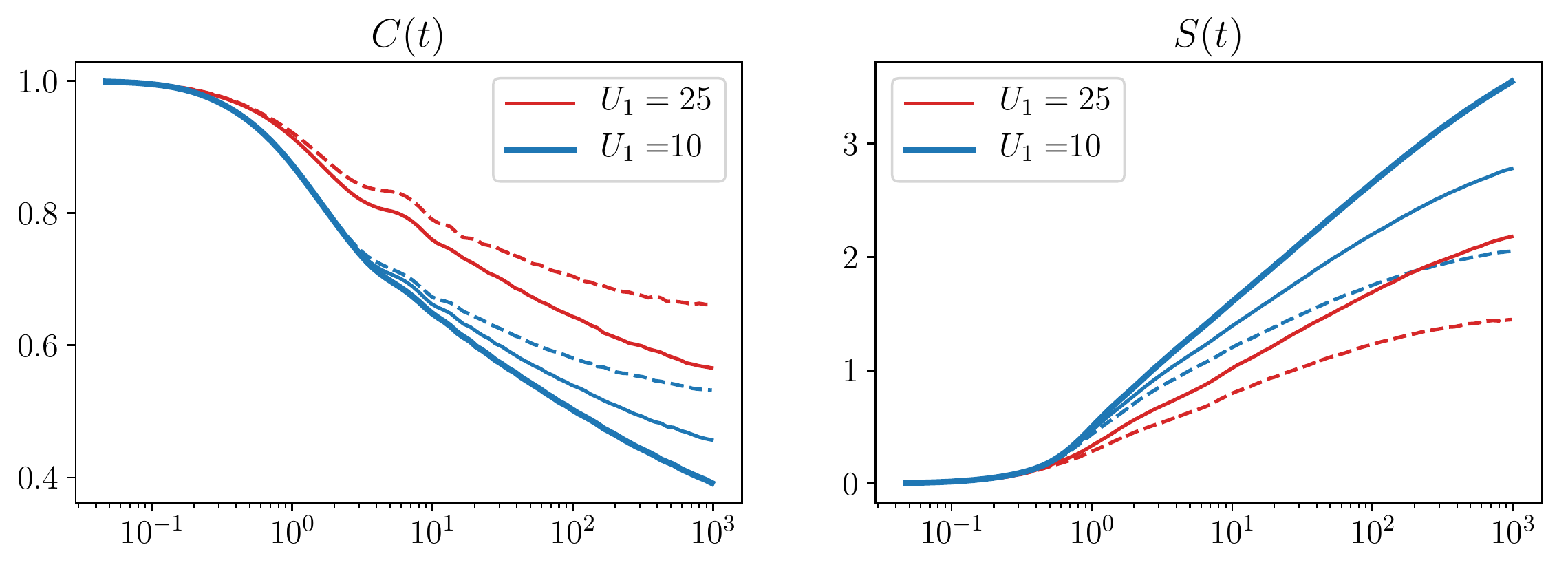}
\end{center}
\caption{\label{nonerg2} {Time evolution of the correlation function, $C(t)$ (left) and of the} entanglement entropy, 
$S(t)$ after a quench {for $W=8$ and two values of $U_1$, and the system initially prepared in} a random Fock state. The dashed, narrow and broad line correspond 
to system size $K=16,18,20$ respectively. Observe that large $U_1$ lead to slower 
entropy growth  and slower decay of the correlation function, in both cases the dependence 
on the system size remains significant. }
\end{figure}

Similar conclusions may be obtained from the time dynamics of the density-density correlation 
function $C(t)$ and of bipartite entanglement entropy $S(t)$ starting from initial random Fock state
as shown in Fig.~\ref{nonerg2}. Large 
$U_1$ leads to slower decay of correlations accompanied by a slower sublogarithmic 
even growth of the entanglement entropy. While for $K>16$ we cannot follow the 
dynamics for sufficiently long times the results indicate a clear saturation of 
$C(t)$ for $U_1=25$ in agreement with mean gap ratio values. 

\section{Time evolution with Chebyshev expansion technique}
\label{cheby}

Time evolution of fermionic systems at half filling with $K=16$ (and smaller) can be obtained easily by 
full exact diagonalization of the Hamiltonian matrix followed by exact calculation of the evolution operator $U(t)$ for 
arbitrary time $t$. 

To deal with larger system sizes we employ the expansion of the evolution operator into series involving Chebyshev polynomials \cite{Weisse06, Bera17}
\begin{equation}
 U(t) \approx \mathrm{e}^{-\mathrm{i}bt} \left( J_0(at) + 2\sum_{k=1}^N (-i)^k J_k(at) T_k \left( \mathcal{H} \right) \right),
 \label{eqcheby}
\end{equation}
where $a=(E_{\rm max} - E_{\rm min})/2$, $b=(E_{\rm max} + E_{\rm min})/2$, the Hamiltonian
is rescaled $\mathcal{H} = \frac{1}{a}(H-b)$ so that spectrum of $\mathcal{H}$ belongs to the $[-1,1]$ interval,
$J_k(t)$ is the Bessel function of the order $k$ and $T_k(x)$ is the Chebyshev polynomial of order $k$.
The number of terms $N$ needed to assure convergence of the expansion  \eqref{eqcheby} to time $t_{\rm max}$ is
$N \approx 2at_{\rm max}$ \cite{Fehske08}.

The time-evolution of the initial state $|\psi_0\rangle$ is given by
\begin{equation}
 |\psi(t) \rangle \approx \mathrm{e}^{-ibt} \sum_{k=0}^N (-i)^k J_k(at)  \left( T_k \left( \mathcal{H}  \right)|\psi_0\rangle \right)
 \label{eqcheby1}
\end{equation}
and reduces to matrix-vector multiplications
\begin{equation}
  T_k \left( \mathcal{H} \right)|\psi_0\rangle = 2 \mathcal{H} T_{k-1}( \mathcal{H})|\psi_0\rangle -
  T_{k-2}( \mathcal{H})|\psi_0\rangle, 
 \label{eqcheby2}
\end{equation}
where the recursion relation satisfied by Chebyshev polynomials was used. In order to get $|\psi(t)\rangle$
we generate iteratively a sequence of $N$ vectors $|\psi_0\rangle, T_{1}|\psi_0\rangle, ...,\, T_{N}|\psi_0\rangle $.
To reach long times of time evolution $t_{\rm max} \approx 10^3$ one needs relatively large $N$ which increases memory consumption.
Therefore we split the time interval $[0, t_{\rm max}]$ into parts $[0, \Delta t ], [\Delta t, 2 \Delta t ], ... $ 
in such a way that $|\psi\left( (n+1)\Delta t \right) \rangle $ can be calculated from the state 
$|\psi\left( n \Delta t  \right)\rangle $
with the expansion \eqref{eqcheby1} involving only a limited number of terms e.g. -- $N\approx 1000$ which allows us to 
obtain time evolution for the system size $K=20$ with memory consumption smaller than $5$GB (performing the matrix-vector 
multiplications in PETSc).

\end{appendix}


\nolinenumbers

\end{document}